\documentclass[aps,prx,twocolumn,epsfig,floats,superscriptaddress]{revtex4}
\usepackage{amsfonts}
\usepackage{color}
\usepackage{braket}
\usepackage{graphicx,epsf,natbib,amsmath,latexsym,amssymb,multirow}
\newcommand{\be}{\begin{equation}}
\newcommand{\ee}{\end{equation}}
\newcommand{\bea}{\begin{eqnarray}}
\newcommand{\eea}{\end{eqnarray}}
\newcommand{\ba}{\begin{array}}
\newcommand{\ea}{\end{array}}

\def \nn {\nonumber}
\newcommand{\eq}[1]{Eq.(\ref{#1})}

\newcommand{\Tr}{\mbox{Tr}}

\newcommand{\e}{\epsilon}

\newcommand{\beq}{\begin{eqnarray}}
\newcommand{\eeq}{\end{eqnarray}}
\newcommand{\bes}{\begin{subequations}}
\newcommand{\ees}{\end{subequations}}

\begin{document}

\title{Decoherence Patterns of Topological Qubits from Majorana Modes}

\author{Shih-Hao Ho}\email{shho@mx.nthu.edu.tw}
\affiliation{Physics Division, National Center for Theoretical Science, Hsinchu, 30013, Taiwan}
\affiliation{Physics Department, National Tsing Hua University, Hsinchu, 30013, Taiwan}

\author{Sung-Po Chao}\email{spchao@gmail.com}
\affiliation{Physics Division, National Center for Theoretical Science, Hsinchu, 30013, Taiwan}
\affiliation{Physics Department, National Tsing Hua University, Hsinchu, 30013, Taiwan}

\author{Chung-Hsien Chou}\email{chouch@mail.ncku.edu.tw}
\affiliation{Department of Physics, National Cheng Kung University, Tainan, 70101, Taiwan}

\author{Feng-Li Lin}\email{linfengli@phy.ntnu.edu.tw} 
\affiliation{Department of Physics, National Taiwan Normal University, Taipei, 11677, Taiwan}
\affiliation{Physics Division, National Center for Theoretical Science, Hsinchu, 30013, Taiwan}

\begin{abstract}

  We investigate the decoherence patterns of topological qubits in contact with the environment by a novel way of deriving the open system dynamics other than the Feynman-Vernon. Each topological qubit is made of two Majorana modes of a 1D Kitaev's chain. These two Majorana modes interact with the environment in an incoherent way which yields peculiar decoherence patterns of the topological qubit.  More specifically, we consider the open system dynamics of the topological qubits which are weakly coupled to the fermionic/bosonic Ohmic-like environments.  We find atypical patterns of quantum decoherence. In contrast to the cases of non-topological qubits for which they always decohere completely in all Ohmic-like environments,  the topological qubits decohere completely in the Ohmic and sub-Ohmic environments but not in the super-Ohmic ones. Moreover, we find that the fermion parities of the topological qubits though cannot prevent the qubit states from decoherence in the sub-Ohmic environments, can prevent from thermalization turning into Gibbs state.  We also study the cases in which each Majorana mode can couple to different Ohmic-like environments and the time dependence of concurrence for two topological qubits.

\end{abstract}
\maketitle
\section{Introduction}

   Topological quantum computation has been a promising scheme of realizing the quantum computer with robust qubits \cite{Nayak:2008zza}. The key ingredient for this scheme is built on the anyonic quasi-particles of topologically ordered systems, which are robust against local perturbations due to its underlying topological nature of quantum orders \cite{topo1}. From quantum entanglement point of view, these topologically ordered states are endowed with the long range entanglement so that the collective anyonic excitations are robust against time-like perturbation \cite{Kitaev:1997wr,LevinWen}. That is, the anyons are long range entangled states so that the local perturbations (i.e., local elementary degrees of freedom in the context of condensed matter or local unitary operations in the context of quantum informations) cannot disentangle the anyonic states.  There are some evidences of finding such intrinsic topologically ordered states in real world, such as Fractional Quantum Hall effect (FQHE). 
However, the anyons with nontrivial anyon statistics as discussed in \cite{anyons-Arovas} in FQHE, which is the key ingredient for realizing universal quantum computation \cite{Nayak:2008zza}, are not yet observed in experiments. Besides, the temperature issue for topological order should be also considered in real world experiments \cite{finiteT}.

   Fortunately, there are new types of topologically ordered states such as topological insulators or superconductors \cite{Fu:2008zzb,HasanKane,Qi2011},  which are easier to be realized physically. For these systems, some excitations are topologically protected as long as some symmetries such as time reversal are preserved. That is, the local perturbations preserving these symmetries cannot disentangle the topological excitations.  Among the topological excitations, the most interesting ones are the Majorana modes localized on the topological defects, which obey the non-abelian anyonic statistics \cite{Mreturn,Arovas,Ivanov}.  The simple mode to realize such  Majorana modes is the Kitaev's chain of 1D spineless p-wave superconductor \cite{Kitaev,Sau,Alicea}. Each on-site fermion $d_i$ can be decomposed into two Majorana modes $\gamma_{2i-1}$ and $\gamma_{2i}$, i.e., $d_i=(\gamma_{2i-1}+i\gamma_{2i})/2$. By appropriately tuning the model, the Majorana modes at the end-points of Kitaev's chain can be dangling without pairing with the other nearby Majorana modes to form usual fermion.  Then, these two far separated end-point Majorana modes can form a topological qubit. The meaning of  ``topological" here is two-fold: one means it is made of Majorana modes $\gamma_1$ and $\gamma_{2N}$ ($N$ denotes the number of regular fermions on the chain) which are topological excitations, and the other means the topological qubit $d_{topo}=(\gamma_1+\gamma_{2N})/2$ itself is non-local, i.e., the two Majorana modes are far separated so that they cannot combine into a usual fermion.  From the quantum information point of view, the topological qubit is EPR-like as it encodes the quantum state in a non-local way. Both features explain its robustness against local perturbations. 
   
   The topological excitations are robust against local perturbations, then one would wonder if the topological qubits are also robust against decoherence when they are considered to be open system by coupling to the non-topological environment. The open system setting is also more realistic when performing the quantum computations. As the quantum informations are carried by physical excitations, the robustness against quantum decoherence implies the robustness against local perturbations but not the other way around. Even the excitations are robust against local perturbations, it is still possible for the quantum informations carried by the topological qubits to leak into the environment. However, as the topological qubit is non-local, the way it interacts with the environment is quite different from the way the usual fermions do, and one would expect the atypical quantum decoherence behaviors. This motivates this work to examine if the topological qubits are robust against quantum decoherence, and their atypical decoherence patterns.  
   
   In fact, it was shown in \cite{Chamon,Budich,Loss} that the topological qubits do decohere by examining either the relaxation time scale or the behavior of the two-time correlator of Majorana modes \cite{superselection}. However, to pin down the decoherence patterns, one should directly study the dynamics of the reduced density matrix of the topological qubits. This is what we do in this paper, and indeed the topological qubits do show atypical decoherence patterns. We take the one spatial dimensional bosonic or fermionic environment, which is universal for $(1+1)$-D conformal field theory such as Luttinger liquid and has Ohmic-like environmental spectral density, i.e.,  $\rho_{spec}(\omega)\propto \omega^Q$ with $Q \ge 0$.  The environment is called Ohmic for $Q=1$, sub-Ohmic for $Q<1$ and super-Ohmic for $Q>1$.  For fermionic coupling this is done by placing a metallic nanowire close to the Majorana endpoint as shown in Fig.\ref{wires}. The Coulomb interaction within the wire can be tuned by choosing different insulating substrate or gating, and the low energy excitation of the wire is deemed as Luttinger liquids. 
   
   We also assume the coupling between Majorana modes and environment to be weak so that Gaussian approximation is good. This also ensures the influence of the Ohmic-like environment to the bulk of the Kitaev's chain is irrelevant so that the robustness of Majorana modes are protected.  We find that the topological qubits decohere completely in the Ohmic and sub-Ohmic environments but not in the super-Ohmic environments.  i.e., they do not relax to the Gibbs state or pointer state.  Thus, in the super-Ohmic environments one may be able to distill the purity or concurrence of the resultant state of topological qubits by appropriate quantum information manipulations. This feature is atypical as compared to the decoherence patterns of the non-topological qubits studied in \cite{Oh,STWu,Hfermion}, for which the local qubits always decohere in all Ohmic-like environments if the probe-environment coupling is weak. Thus, we conclude that the topological nature does protect the topological qubits from decoherence to some extent.

   We organize our paper as follows. In the next section we briefly introduce the setup of topological qubits made by Majorana modes interacting with the environment via coupling to either fermionic or bosonic operators. We then develop our interaction picture formalism for deriving the open system dynamics of the topological qubits. In section \ref{secDQ} we study the decoherence patterns of single and two topological qubits weakly coupled to the Ohmic-like  environments. We obtain the explicit form the reduced density matrix for the single topological qubit to see the decoherence patterns. For two topological qubits, various results about the robustness against complete decoherence are fully exposed both analytically and numerically. Finally, we conclude our paper in section \ref{secCon}.  Various technical details about the real time Green functions and explicit form of reduced density matrices are given in Appendices.
   
\section{Dynamics of open system for Majorana qubits}
   In this section we consider an open system of topological qubits made of Majorana modes and study the dynamics of its reduced density matrix. The peculiar features of Majorana modes such as obeying the Clifford algebra make the consideration of its decoherence far simpler than the usual simple harmonic oscillator probe. As we show below the reduced dynamics of this open system by integrating out the environment can be obtained in a closed form when formulating the formalism in the interaction picture. This is in contrast to the usual Feynman-Vernon formulation for the non-topological qubits \cite{WMZhang,Hfermion}, for which one needs to numerically solve involved Langevin-like equation.

\subsection{Open system for Majorana modes}

    The system considered in this paper is described by the following Hamiltonian
\beq \label{E01}
H=\hat{H}_0+\hat{V}=\hat{H}_{\mathcal{P}}+\hat{H}_{\mathcal{E}}+\hat{V}
\eeq
where $\hat{H}_0$ is the free Hamiltonian consisting of $\hat{H}_{\mathcal{P}}$ for the probe $\mathcal{P}$ and $\hat{H}_{\mathcal{E}}$ for the environment $\mathcal{E}$, and $\hat{V}$ is the interaction between $\mathcal{P}$ and $\mathcal{E}$. We also assume $[\hat{H}_{\mathcal{P}}, \hat{H}_{\mathcal{E}}]=0$.

\begin{figure}
\includegraphics[width=.8\columnwidth]{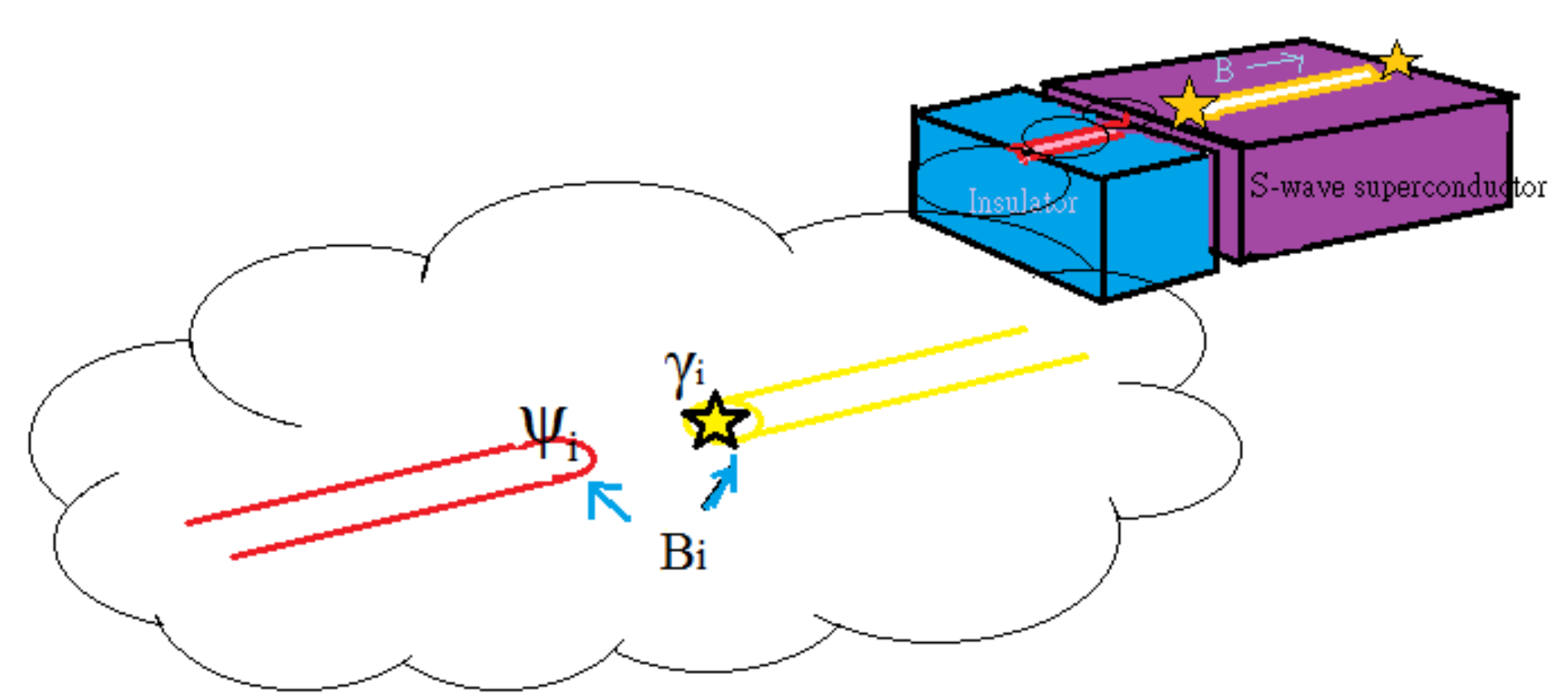}
\caption{Schematic diagram for Majorana modes coupled to fermionic environments: Majorana modes are generated at end points of some nanowire with strong spin-orbit interaction, placed on top of a s-wave superconductor, and an external magnetic field is applied along wire axis direction. Each Majorana mode (shown as a gold star) is coupled to a metallic nanowire via a tunnel junction (only one side is shown for simplification), with tunneling strength $B_i$ controllable by an external gate voltage. The effective Coulomb interaction of the metallic wire can be tuned by placing at different substrates (shown as blue region) which modifies its dielectric constants.}
\label{wires}
\end{figure}

 Here the probe consists of a bunch of Majorana modes localized at the ends of some quantum wires \cite{Kitaev}, see also \cite{expts,Fisher} for experimental proposals. We denote these localized Majorana modes as $\gamma_a$, with $a=1,2,\cdots$, which have the following properties: 
\be\label{majCond}
\gamma_a^{\dagger}=\gamma_a\;, \qquad  \{\gamma_a, \gamma_b\}=2\delta_{ab}. 
\ee
On the other hand, the dynamics of the environment is dictated by $\hat{H}_{\mathcal{E}}$ whose elementary constituents can be thought as electrons or holes. Due to the aforementioned peculiar properties of $\gamma_a$'s, the generic interaction Hamiltonian takes the form
\be\label{majV}
\hat{V}=\sum_{a} B_a \gamma_a \mathcal{O}_a + \sum_{a>b} B_{ab} \gamma_a \gamma_b \mathcal{O}_{ab} +\cdots
\ee
where $B_a$'s and $B_{ab}$'s are real coupling constants, and $\mathcal{O}_a$'s and $\mathcal{O}_{ab}$'s are composite operators of electrons' creation and annihilation operators $\psi^{\dagger}_{\alpha}$'s and $\psi_{\alpha}$'s with $\alpha=a,b$ labels the electrons' bath and spin indices. For the tunneling junction shown in Fig.\ref{wires}, $\mathcal{O}_a=\psi^{\dagger}_a-\psi_a$. The $\cdots$ denotes the higher order terms involving more number of Majorana modes, which however are not considered in this paper. Moreover, note that
\be\label{antihermi}
\mathcal{O}^{\dagger}_a = -\mathcal{O}_a\;, \qquad \mathcal{O}^{\dagger}_{ab}=-\mathcal{O}_{ab}\;.
\ee
which follow from the Hermitian condition of the full Hamiltonian. 

\begin{figure}
\includegraphics[width=0.4\columnwidth]{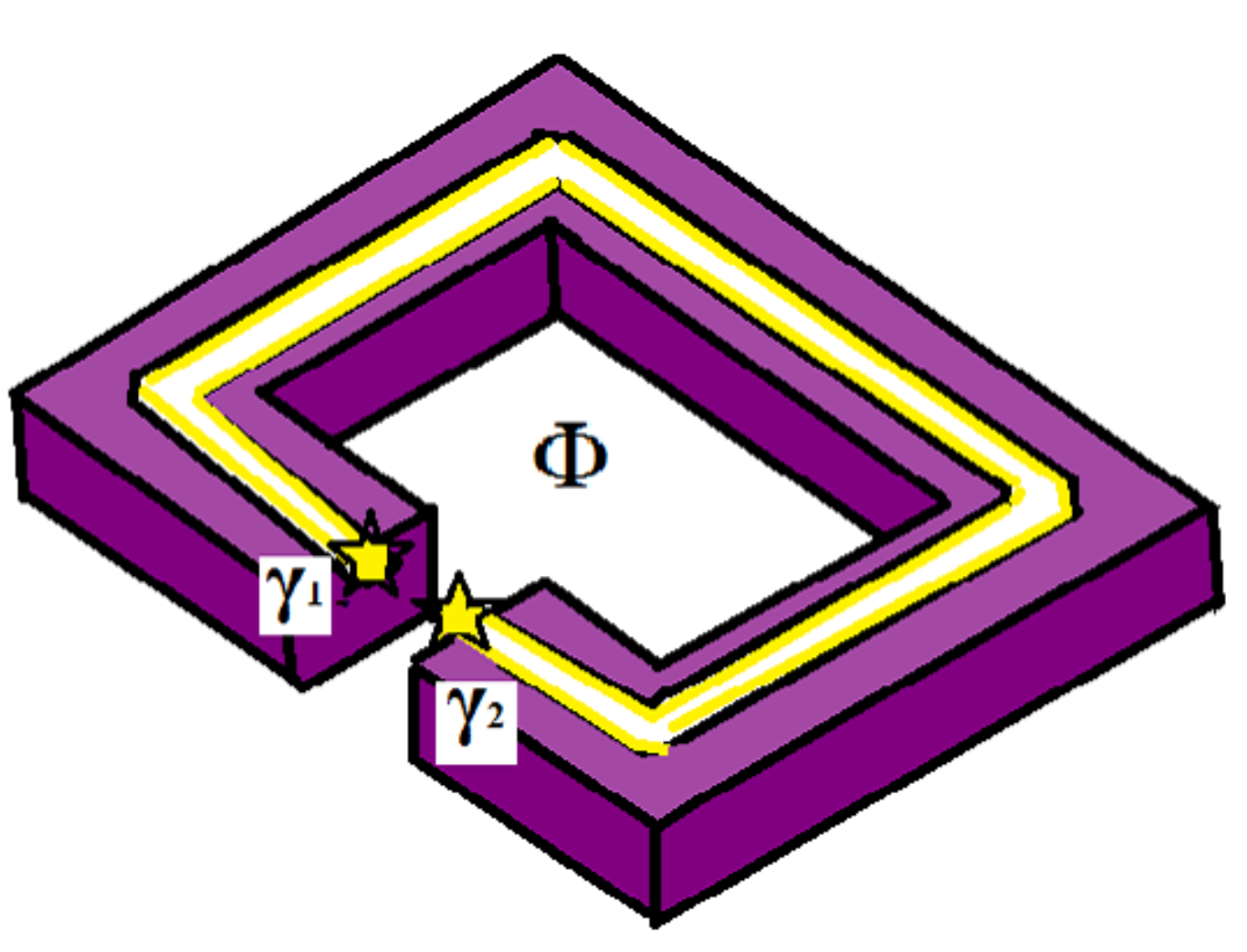}
\caption{Schematic diagram for two Majorana modes coupled with a bosonic environment. The two Majorana modes are located at the endpoints of a ring with a small gap in between. The pair of the Majorana modes $\gamma_1$, $\gamma_2$ interact with some effective environmental bosonic operator locally. The frequency dependence in the bosonic environment in principle can be generated by an external time dependent magnetic flux $\Phi$, similar to the proposal in Ref.~\onlinecite{Hu}.}
\label{ringT}
\end{figure}

As the Majorana modes could be spatially separated, the second term in \eq{majV} could be non-local, i.e., $\gamma_a \gamma_b \mathcal{O}_{ab}:=\gamma_a(\bf{x}) \mathcal{O}_{ab}(\bf{x},\bf{y}) \gamma_b(\bf{y})$.  For simplicity, in this work we only consider the local interactions. In such a case, the locality for the two localized Majorana modes can be arranged as either in Fig.\ref{wires} for neighboring wires or Fig.\ref{ringT} for a ring with a small gap.   

 The density matrix $\hat{\rho}(t)$ for the whole system in the Schr\"odingier picture evolves as
\beq \label{E02}
\hat{\rho}(t)=\Ket{\psi(t}\Bra{\psi(t)} = e^{-i\hat{H}t}\; \hat{\rho}_0 \; e^{ i\hat{H}t}
\eeq
where $\hat{\rho}_0 \equiv \hat{\rho}(t=0)$. We assume the direct product structure for $\hat{\rho}_0$, i.e.,
\be\label{factorrho}
\hat{\rho}_0=\hat{\rho}_{\mathcal{P}}  \otimes \hat{\rho}_{\mathcal{E}}\;.
\ee

  In general, $\hat{V}$ mixes probe's and environment's degrees of freedom so that $[\hat{H}_0, \hat{V}] \ne 0$. This prevents further simplification when evaluating the reduced density matrix for probe $\mathcal{P}$, i.e.,
\be \label{rdm-1}
\hat{\rho}_r(t)= \mbox{Tr}_{\mathcal{E}}  \;  e^{-i\hat{H}t}\; \hat{\rho}_0 \; e^{ i\hat{H}t}
\ee
where $\mbox{Tr}_{\mathcal{E}}$ is to take trace over the Hilbert space of environment $\mathcal{E}$. We rewrite \eq{rdm-1} as 
\beq\label{rdm-2}
\langle i | \hat{\rho}_r(t)|  j \rangle = \sum_{m_+,m_-, K,L_+,L_-} \langle i,K| e^{-i\hat{H}t}| m_+,L_+\rangle\ \\\nn \times \langle m_+,L_+| \hat{\rho}_0 | m_-,L_-\rangle\; \langle m_-, L_-| e^{i\hat{H}t}| j, K\rangle
\eeq
where the lower case letters such as $i,j,m_{\pm}$ are the labels for the probe's state space, and the upper case ones such as $K,L_{\pm}$ are the ones for the environment's state space. Though we use the notation for the discrete labels, the generalization to the continuum is straightforward. 

The first and third factors in \eq{rdm-2} can be understood as the Schwinger-Keldysh Green functions \cite{Schwinger:1960qe, Keldysh:1964ud} on the forward (labeled by the subscript ``+") and backward (labeled by the subscript ``-") real-time Keldysh contour \cite{Keldysh:1964ud}, respectively.  Due to the fact that $[\hat{H}_0, \hat{V}] \ne 0$, it is usually difficult to further simplify these Green functions. The standard practice in deriving the dynamics of the reduced density matrix is to adopt Feynman-Vernon formalism \cite{Feynman:1963fq,Caldeira:1982iu,Grabert:1988yt} by integrating out the environment in the path-integral formulation and obtain a real time effective theory (the so-called ``influence functional") for the probe.

\subsection{Interaction picture formulation}

     The Feynman-Vernon formalism usually involves solving Langevin-like equation with complicated non-local kernel in order to explicitly obtaining $\hat{\rho}_r(t)$. This usually needs careful numerical computations. Here we show that for probe composed by Majorana modes, even with $[\hat{H}_0, \hat{V}] \ne 0$, the calculations of obtaining $\hat{\rho}_r(t)$ in the formulation of interaction picture are greatly simplified, with no need of introducing path integral and solving Langevin-like equation. We elaborate this formulation as follows.
     
      The full density matrix in the interaction picture is
\bes \label{E03}
\beq \label{E03-1}
\hat{\rho}_I(t) &\equiv& e^{i \hat{H}_0 t} \hat{\rho}(t) e^{-i \hat{H}_0 t} \\\nn &=&  e^{i \hat{H}_0 t} e^{-i\hat{H}t} \hat{\rho}_0 e^{i\hat{H}t} e^{-i \hat{H}_0 t} =U(t) \hat{\rho}_0 U^{\dagger}(t)
\eeq
where the evolution operator $U(t) \equiv e^{i \hat{H}_0 t} e^{-i\hat{H}t}$. The evolution operator satisfies the Schr\"odinger equation 
\be
\label{E03-2}
i \frac{d}{dt} U(t) = \hat{V}_I(t) U(t), 
\ee
where 
\be
\hat{V}_I(t) \equiv e^{i\hat{H}_0 t}\; \hat{V}\; e^{-i\hat{H}_0 t}\;.
\ee
For the interaction \eq{majV} considered here, we have 
\be
\hat{V}_I(t) = \sum_{a} B_a \gamma_a(t) \mathcal{O}_a(t) + \sum_{a<b} B_{ab} \gamma_a\gamma_b(t) \mathcal{O}_{ab}(t)
\ee
where
\be
\mathcal{O}_{\mathcal{P}}(t)=e^{-i H_{\mathcal{P}} t} \mathcal{O}_{\mathcal{P}} e^{i H_{\mathcal{P}} t}\;,\mathcal{O}_{\mathcal{E}}(t)=e^{-i H_{\mathcal{E}} t} \mathcal{O}_{\mathcal{E}} e^{i H_{\mathcal{E}} t}
\ee
with $\mathcal{O}_{\mathcal{P}}=\gamma_a$ or $\gamma_a \gamma_b$, and $\mathcal{O}_{\mathcal{E}}=\mathcal{O}_a$ or $\mathcal{O}_{ab}$.

We then solve \eq{E03-2} to arrive
\be
\label{E03-3}
U(t) = \textrm{T} e^{-i \int^t d\tau \hat{V}_I(\tau)}.
\ee
Here $\textrm{T}$ is the time-ordering operator on the forward Keldysh contour. We also denote the time-ordering on the backward Keldysh contour as $\tilde{\textrm{T}}$ such that
\be
U^{\dagger}(t) = \tilde{\textrm{T}} e^{i \int^t d\tau \hat{V}_I(\tau)}.
\ee

     It is easy to check that  $\hat{\rho}_I(t)$  satisfies the following equation of motion, i.e., Schr\"odingier equation for the density matrix, 
\beq \label{E03-4}
\frac{d}{dt} \hat{\rho}_I(t) = -i [\hat{V}_I(t), \hat{\rho}_I(t)]\;.
\eeq
\ees
      
      Based on the interaction picture, it is easy to see that the reduced density matrix for probe $\mathcal{P}$ can be further simplified as follows
\bea\nn
\hat{\rho}_r(t)&&=e^{-i \hat{H}_{\mathcal{P}} t} \left( \mbox{Tr}_{\mathcal{E}} e^{-i\hat{H}_{\mathcal{E}} t} \hat{\rho}_I(t) e^{i \hat{H}_{\mathcal{E}} t} \right) e^{ i \hat{H}_{\mathcal{P}} t}\;, \\\nn
&&=e^{-i \hat{H}_{\mathcal{P}} t} \left( \mbox{Tr}_{\mathcal{E}}  \hat{\rho}_I(t)   \right) e^{ i \hat{H}_{\mathcal{P}} t}\;, \\
&&=e^{-i \hat{H}_{\mathcal{P}} t} \left( \mbox{Tr}_{\mathcal{E}} U(t) \hat{\rho}_0  U^{\dagger}(t) \right) e^{ i \hat{H}_{\mathcal{P}} t}\;. \label{rhointer}
\eea
In the first line we have used the fact $[\hat{H}_{\mathcal{P}}, \hat{H}_{\mathcal{E}}]=0$, and in the second line we simply use the cyclic property of trace operation and $e^{-i\hat{H}_{\mathcal{E}} t} e^{i \hat{H}_{\mathcal{E}} t} =1 $.  In the final expression, the $\hat{H}_{\mathcal{E}}$ dependence is implicitly hidden in $\mathcal{O}_{\mathcal{E}}(t)$.  

\subsection{Reduced density matrix for Majorana probe}

   As these Majorana modes used for the qubits are zero energy modes, we have
\be
\hat{H}_{\mathcal{P}}=0
\ee
 and thus $\mathcal{O}_{\mathcal{P}}(t)=\mathcal{O}_{\mathcal{P}}$, i.e., $\gamma_a(t)=\gamma_a$, etc.

To obtain the simple analytic reduced dynamics, we consider the interaction Hamiltonian in which the individual terms in \eq{majV} commute with each other. For example, we can not have both $\gamma_1 \mathcal{O}_1$ and $\gamma_1 \gamma_2 \mathcal{O}_{12}$, or both $\gamma_1\gamma_2 \mathcal{O}_{12}$ and $\gamma_2\gamma_3 \mathcal{O}_{23}$. Otherwise, we would not have the closed form for the reduced dynamics. Though this constraint looks artificial just for simplifying the derivation, it can be naturally realized  once the shape of the nanowire is given as the Majorana modes are confined to the endpoints of the wire. For example, as $\gamma_1$ and $\gamma_2$ are located at two ends of the nanowire, we can only have $\gamma_i \mathcal{O}_i$ but not $\gamma_1 \gamma_2 \mathcal{O}_{12}$ for a straight wire. On the other hand, for a ring wire with an extremely narrow gap as in Fig. \ref{ringT} it is natural to just have $\gamma_1 \gamma_2 \mathcal{O}_{12}$ but not $\gamma_i \mathcal{O}_i$.

Given the above constraint the evolution operator $U(t)$ can be decomposed as:
\bea\label{factorU}
&&U(t):= \textrm{T} \; \Pi_M e^{-i \Gamma_M  {\bf O}_M(t) }\\\nn
&&=\textrm{T} \; \Pi_a e^{-i B_a \gamma_a \int^t d\tau \mathcal{O}_a(\tau) } \; \Pi_{a<b} \; e^{-i B_{ab} \gamma_a \gamma_b  \int^t d\tau \mathcal{O}_{ab}(\tau)}\;.
\eea
Here we combine the set of the operator indices $a$ and $ab$ into a unified symbol $M$ so that $\Gamma_M=\gamma_a$ or $\gamma_a \gamma_b$, and so on. Moreover, for simplicity we have denoted $B_M \int^t d\tau \mathcal{O}_M (\tau)$ by ${\bf O}_M(t)$ with the time-ordering being taken care. 

   Using the properties given in \eq{majCond} of $\gamma_a$'s, each factor in \eq{factorU} is further simplified, viz.,
\be\label{oaga}
\textrm{T}\; e^{-i \Gamma_M {\bf O}_M(t) } = \textrm{T}\; [ \cosh {\bf O}_M(t) - i \Gamma_M \sinh {\bf O}_M(t)]\;.
\ee
Note that in the above we have used the fact that ${\bf O}_a$'s are fermionic operators and ${\bf O}_{ab}$'s are the bosonic ones. 
Thus the time ordered evolution matrix is
\be
U(t)=\textrm{T}\; \Pi_M [\cosh {\bf O}_M(t) - i \Gamma_M \sinh {\bf O}_M(t)]\;.
\ee
Similarly, the anti time-ordered one is
\be
U^{\dagger}(t)= \tilde{\textrm{T}} \; \Pi_M [\cosh {\bf O}_M(t) + i \Gamma_M \sinh {\bf O}_M(t)]\;.
\ee

    We further simplify \eq{rhointer} by using the fact that $\Tr_{\mathcal{E}}(\hat{\rho}_{\mathcal{E}} \mathcal{O}_{\mathcal{E}})=0$ if $\mathcal{O}_{\mathcal{E}}$ is a fermionic operator of the environmental theory, i.e.,
\be\label{nofvev}
\langle \mathcal{O}_{\mathcal{E}} \rangle_{\mathcal{E}}=0 \quad \mbox{if $\mathcal{O}_{\mathcal{E}}$ is fermionic.}
\ee   
 Thus, the reduced density matrix should take the form
\beq\nn
\hat{\rho}_r(t)&=&\sum_{\{m\},\{m'\}} e^{i\phi_{m,m'}} [ \Pi_{\{m\}}\mathcal{G}_m \hat{\rho}_{\mathcal{P}} \Pi_{\{m'\}} \mathcal{G}_{m'}]_+  \\ \label{rhofinal} & \times&\langle [\tilde{\textrm{T}}\;\Pi_{\{m'\}} \mathcal{H}_{m'}(t) \; \textrm{T} \; \Pi_{\{m\}} \mathcal{H}_m(t) ]_+\rangle_{\mathcal{E}}
\eeq
where $[\cdots]_+$ denotes the operator $\cdots$ is of even fermion parity,  $\mathcal{H}_m(t)$ is either $\cosh {\bf O}_M(t)$ or $\sinh {\bf O}_M(t)$,  and $\mathcal{G}_m$ in $\Pi_{\{m\}}\mathcal{G}_m$ is 1 if $\mathcal{H}_m(t)=\cosh {\bf O}_M(t)$, otherwise it is $\Gamma_M$. The phase factor $e^{i\phi_{m,m'}}=\pm 1, \pm i$ is determined by the relative ordering of the fermionic operators, and $\sum_{\{m\},\{m'\}}$ runs over all possible sets of even fermion parity terms in the binomial expansion of $U\hat{\rho}_0 U^{\dagger}$.  Here, we have used the short-handed notation for the the real-time Green function, i.e.,
\beq\nn
&&\langle [\tilde{\textrm{T}}\;\Pi_{\{m'\}} \mathcal{H}_{m'}(t) \; \textrm{T} \; \Pi_{\{m\}} \mathcal{H}_m(t) ]_+\rangle_{\mathcal{E}}\\&&:= \Tr_{\mathcal{E}}  [ \textrm{T} \; \Pi_{\{m\}} \mathcal{H}_m(t) \hat{\rho}_{\mathcal{E}} \tilde{\textrm{T}}\;\Pi_{\{m'\}} \mathcal{H}_{m'}(t) ]_+ \;.
\eeq

   Due to the locality of the Majorana modes, we assume the environments for different channels are independent, i.e.,
\be\label{indp-cc}
\langle \mathcal{O}_{\mathcal{E}} \tilde{\mathcal{O}}_{\mathcal{E}} \rangle_{\mathcal{E}}=0, \qquad \mbox{if} \quad \mathcal{O}_{\mathcal{E}}\ne \tilde{\mathcal{O}}_{\mathcal{E}}\;.
\ee
In such a case, \eq{rhofinal} is further simplified as
\beq\nn 
\hat{\rho}_r(t)&=& \sum_{\{m\}} (-1)^{{1\over 2} m_F(m_F-1)}  [ \Pi_{\{m\}}\mathcal{G}_m  \hat{\rho}_{\mathcal{P}} \Pi_{\{m\}}\mathcal{G}_m ]_+ \\\label{rhofinal-s} &\times&  [\Pi_{\{m\}}  \langle \tilde{\textrm{T}}\; \mathcal{H}_{m}(t) \; \textrm{T} \;  \mathcal{H}_m(t) \rangle_{\mathcal{E}}]_+\;,
\eeq 
where $m_F$ is the number of fermionic operators in $\Pi_{\{m\}}  \mathcal{H}_m$, i.e., $\cosh \mathcal{O}_a$'s, $\cosh \mathcal{O}_{ab}$'s and $\sinh \mathcal{O}_{ab}$'s are bosonic but $\sinh \mathcal{O}_a$'s are fermionic.  Here $\sum_{\{m\}}$ runs over all the binomial terms of $\Pi_M (\cosh {\bf O}_M +  \sinh {\bf O}_M)$ for $\Pi_{\{m\}} \mathcal{H}_{m}$, and $\mathcal{G}_m$ in $\Pi_{\{m\}}\mathcal{G}_m$ is 1 if $\mathcal{H}_m(t)=\cosh {\bf O}_M(t)$, otherwise it is $\Gamma_M$. Again, due to \eq{nofvev} those non-vanishing terms are of even fermion parity.

   Before we consider the specific cases to obtain explicit dynamics of $\hat{\rho}_r(t)$, three important remarks about arriving the form of \eq{rhofinal-s} should be mentioned:
\begin{itemize}   
   \item As both the probe and the environment factors in \eq{rhofinal-s} are of even fermion parity, and thus bosonic and commuting with each other,  the total Hilbert space can be casted into the form of direct products of the one for probe and the other one for environment, i.e., \eq{factorrho}.  Thus, we can evaluate the first factor of the Majorana modes in terms of the finite dimensional representation of the Clifford algebra $\{ \gamma_a, \gamma_b \}=2\delta_{ab}$.  
   
   \item Even $\mathcal{O}_a$ is a fermionic operator, we should treat it as the bosonic operator when evaluating its corresponding real-time Green function in \eq{rhofinal-s} \cite{Chamon}. This can be understood from the fact that $\gamma_a$ always accompanies  $\mathcal{O}_a$ as indicated in \eq{oaga}, so the minus sign arising from switching of $\mathcal{O}_a$'s as required by the time-ordering is canceled by the analogous minus sign for $\gamma_a$. The latter minus sign is implicit as we set $\hat{H}_{\mathcal{P}}=0$. To make this explicit, we assume $\hat{H}_{\mathcal{P}}\ne 0$, then
\beq\nn
&&\langle \textrm{T}_C\; \mathcal{O}_a(t) \gamma_a(t)\gamma_a(t') \mathcal{O}_a(t')  \rangle_{\mathcal{E}}\\\nn &&=\gamma_a(t)\gamma_a(t')\langle  \mathcal{O}_a(t) \mathcal{O}_b(t')  \rangle_{\mathcal{E}} \Theta(t-t') \\\nn&&+\gamma_a(t') \gamma_a(t)\langle  \mathcal{O}_a(t') \mathcal{O}_a(t)   \rangle_{\mathcal{E}} \Theta(t'-t)
\eeq
where $\textrm{T}_C$ is the time-ordering for the Keldysh contour.  For $\hat{H}_{\mathcal{P}}=0$ we then have $\gamma_a(t)=\gamma_a$. Thus $\gamma_a(t)\gamma_a(t')=1=\gamma_a(t') \gamma_a(t)$ and        
\beq\nn
&&\langle \textrm{T}_C\; \mathcal{O}_a(t) \gamma_a\gamma_a\mathcal{O}_a(t')  \rangle_{\mathcal{E}}\\\nn &&= \langle  \mathcal{O}_a(t) \mathcal{O}_b(t')  \rangle_{\mathcal{E}} \Theta(t-t')+ \langle  \mathcal{O}_a(t') \mathcal{O}_a(t)   \rangle_{\mathcal{E}} \Theta(t'-t)\;.
\eeq           
       
We call the above Green functions the ``Majorana-dressed Green functions", and they are bosonic even the operators $\mathcal{O}_a$ are fermionic. Moreover, the Majorana-dressed Green functions for the operators $\mathcal{O}_{ab}$'s are the same as the undressed ones. Later we denote the Majorana-dressed Green function as $\overline{G}(t,t')$ or $\overline{G}(\omega)$.

   \item In contrast to the usual Lindblad formalism, for example, by applying it to the 1D  and 2D  topological insulators \cite{Lindblad-1,Lindblad-2}, our formalism is different in two aspects: (i) We obtain the reduced density matrix \eq{factorrho} directly by utilizing the Clifford algebra of Majorana modes without the need of solving the master equation. (ii) We resum the diagrams in Gaussian form as opposed to the Born approximation usually used in the Lindblad formalism, in which the perturbation from the probes are considered only up to second order. Furthermore we do not assume Markov approximations. As shown later our results do have some non-Markovian behaviors.
         
\end{itemize}

\subsection{Environmental influence functional}   \label{section3a}

       The second factor of \eq{rhofinal-s} is the environmental influence functional to the dynamics of the probe qubits. They are nothing but the product of real-time (bosonic) Green functions of the even-parity sectors.  Note that the influence functional completely determines the dynamics of the probe as we set $\hat{H}_{\mathcal{P}}=0$. From \eq{indp-cc} we have the following bosonic real-time correlation functions appearing in \eq{rhofinal-s}:
\be \label{CCSS}
\langle \tilde{\textrm{T}}\mathcal{H}_m(t) \textrm{T} \mathcal{H}_m(t) \rangle_{\mathcal{E}}
\ee
where $\mathcal{H}_m$ can be either $\cosh {\bf O}_M(t)$ or $\sinh {\bf O}_M(t)$ with $M=a$ or $ab$, and also
\be\label{even-G}
 \langle \tilde{\textrm{T}}\cosh {\bf O}_{ab}(t) \textrm{T} \sinh {\bf O}_{ab}(t) \rangle_{\mathcal{E}}\;,  \quad  \langle \tilde{\textrm{T}} \sinh {\bf O}_{ab}(t) \textrm{T} \cosh {\bf O}_{ab}(t) \rangle_{\mathcal{E}}\;. 
\ee

   In practical it is difficult to evaluate the real-time correlation functions in \eq{CCSS} in an exact way. However, in this paper we assume all the coupling constants, i.e., $B_a$'s and $B_{ab}$'s are weak so that we first expand these correlation functions up to second order in the coupling constants and then perform the appropriate re-exponentiation \cite{Hu:1991di,Boyanovsky:2004dj} to approximate the original correlation functions in \eq{CCSS}, especially their long time behaviors. In this way, the results are expressed in term of  Schwinger-Keldysh Green functions, see Appendix  \ref{A0} for the detailed definitions. 

  The aforementioned  re-exponentiation procedure is equivalent to resuming the one-particle irreducible diagrams, and should be performed with great care \cite{reexp-fn}, see Appendix \ref{reexp} for more detailed discussions. to capture the precise long-time behavior, at least qualitatively. 
 In the end, we obtain the following results:
\bea\nn \label{cc1}
&&\langle \tilde{\textrm{T}} \cosh{\bf O}_M(t) \textrm{T} \cosh{\bf O}_M(t) \rangle_{\mathcal{E}}\\\label{ccG}&&\approx {1\over 2}(e^{2 B_M^2 \int^t d\tau \int^t d\tau' \overline{G}_{M,sym}(\tau-\tau') }+1)\;,
\\ \nn \label{ss1}
&&\langle \tilde{\textrm{T}} \sinh{\bf O}_M(t) \textrm{T} \sinh{\bf O}_M(t) \rangle_{\mathcal{E}}\\\label{ssG}&& \approx {1\over 2}( e^{2 B_M^2 \int^t d\tau \int^t d\tau' \overline{G}_{M,sym}(\tau-\tau') }-1)\;.
\eea
 Here $\overline{G}_{M,sym}$ is the Majorana-dressed symmetric Green function as defined in \eq{25}. Note that the retarded Green function does not show up in \eq{cc1} and \eq{ss1}.  

   On the other hand, the correlation functions in \eq{even-G} are related to one-point function after the expanding up to the second order and re-exponentiating (resuming the tadpole diagrams), e.g.,
\be\label{condensate}
\langle \tilde{\textrm{T}}\cosh {\bf O}_{ab}(t) \textrm{T} \sinh {\bf O}_{ab}(t) \rangle_{\mathcal{E}}\approx e^{B_M \int^t d\tau \langle \mathcal{O}_{ab}(\tau) \rangle_{\mathcal{E}}}-1\;,
\ee
and thus vanishes if there is no condensate of $\mathcal{O}_{ab}$. For nonzero, time independent, condensate such as the simple mean field result of superconducting order parameter, \eq{condensate} does not vanish and the obtained reduced density matrix shows oscillating off-diagonal components. Since the bosonic condensate alone will not cause decoherence of the topological qubits, we assume no condensate in this paper to explore the physics beyond its effect.
     
\subsection{Implication for the dissipation-less Majorana modes}
     
     From the above results, we see that the time dependence of the reduced density matrix is controlled by the double integral of the symmetric Green functions and the scalar condensates. This is quite different from the usual influence functional for the non-Majorana probe, which is in general involved with both the retarded and symmetric Green functions \cite{Caldeira:1982iu,Grabert:1988yt,Hu:1991di,Son:2009vu,Ho:2013rra}, i.e., the influence functional taking the form
\be
e^{- g^2 \int_{t_i}^{t_f} d\tau  \int_{t_i}^{t_f} d\tau' 
\; [ \Delta(\tau) G_{\textrm{R}}(\tau-\tau') \Sigma(\tau')-{i\over 2} \Delta(\tau) G_{\textrm{sym}}(\tau-\tau') \Delta(\tau') ]}
\ee
where $\Sigma(\tau)$ and $\Delta(\tau)$ are the center of mass and relative coordinates in the so-called ``ra" basis \cite{Caldeira:1982iu,KCChou,Son:2002sd}. 
     
     Especially, the KMS relation between retarded and symmetric Green functions and their appearance in the influence functional of Feynman-Vernon yield the fluctuation-dissipation theorem for the Brownian motion of the probe.  More specifically, the Langevin equation derived from the influence functional for the usual Brownian particle takes the form
\be\label{langevin}
\ddot{\Sigma}+\omega^2 \Sigma + g^2 \int^t d\tau G_R(t-\tau) \Sigma(\tau) = \xi(t)
\ee
with 
\be
\langle \xi(t)\xi(t')\rangle_{\mathcal{E}} = g^2 G_{sym}(t-t')\;.
\ee
Here, the dissipation term is controlled by the retarded Green function and the fluctuation one is controlled by the symmetric Green function. Thus the KMS relation \eq{KMS} yields the fluctuation-dissipation theorem.  Furthermore, the kernel for evolving probe's reduced density matrix is related to the solutions of the Langevin equation \eq{langevin}, in which the dissipation kernel term $\int^t d\tau G_R(t-\tau) \Sigma(\tau)$ plays an essential role in the reduced dynamics for quantum decoherence, for example see \cite{Caldeira:1982iu,Hu:1991di,Su:1987pi,Ho:2013rra}.

       Our method is different from the Feynman-Vernon's so that our ``influence functional" may not be exactly the same as the Feynman-Vernon's. However, due to the physical similarity between these twos, it is tempting to say that the absence of the retarded Green function in the influence functional for the Majorana mode implies that its dynamics is dissipation-less. 
       
      The dissipation-less feature of Majorana modes are expected as its transport is related to anomaly transport \cite{Ludwig}, which is shown to be dissipation-less hydro-dynamically, i.e., there is no generation of entropy \cite{Haehl:2013hoa}.  Intuitively, if the environment is not topologically ordered so that it cannot produce a dangling Majorana mode to combine with the end-point one into an electron, then the end-point Majorana mode is robust. However, there is still a possibility that the environment's electron deconfines into a pair of Majorana modes, one of them combine with the end-point Majorana mode to turn into an end-point electron, and the other one leaks into the environment. This is in fact captured by the interaction Hamiltonian $\hat{V}$ of \eq{majV}, which implies that the end-point Majorana mode can turn into an electron mode. 
      
       For the whole system the interaction Hamiltonian still preserves the Majorana mode number, we can just think that the end-point Majorana mode just moves into the bulk. However, from the point of view of the open system the end-point Majorana mode just dissipates away or thermalizes. Despite that, our resultant ``influence functional" with the absence of retarded Green function suggests that the dissipation-less feature of the Majorana modes may still preserve to some extent even under the effect of interaction Hamiltonian. To pin down more specifically the dissipation-less feature for the Majorana modes and understand the dissipation-fluctuation theorem in sense of Langevin, one may need to derive Feynman-Vernon's influence functional. We will not consider this issue further in this work. Instead, it is interesting to ask if such a possible dissipation-less feature also implies robustness against decoherence or not. 
     
     Recall the fact that the Fourier transform of the symmetric Green function encodes the spectral density of the corresponding channel for fluctuation and dissipation. However, the dissipation ability of the effective carriers for decoherence is given by $\int^t d\tau \int^t d\tau' \overline{G}_{M,sym}(\tau-\tau')$ as suggested in \eq{cc1} and \eq{ss1}. Thus, there is a possibility that the effective carriers are not so efficient to carry away the quantum information of the topological qubits as the spectral density implies. We see this is indeed the case in the next section.

\section{Decoherence Patterns of Topological Qubits}\label{secDQ}
  
    Now we are ready to apply the above formal results for studying the patterns of decoherence of the topological qubits in some environments. As the influence functional due to the environment (taking the form of $\int^t d\tau \int^t d\tau' \overline{G}_{M,sym}(\tau-\tau')$ in our case) affects the decoherence pattern, one should specify the dynamics of the environment to obtain the concrete results. For the fermionic environment we consider the setup of quantum wire \cite{expts,Fisher} tunnel coupled to the superconducting wires as shown in Fig. \ref{wires}. We take the helical Luttinger wire \cite{SungPo1}, which could be realized as the edge state of some two dimensional topological insulator \cite{Zhang}, as a special example, but the generalization to other types of Luttinger liquids \cite{Kane} is straightforward. For the bosonic environment we take the ring structure \cite{Hu}, as shown in Fig. \ref{ringT}, with external magnetic flux $\Phi$ controlling different frequency modes of bosonic couplings. In general the Majorana modes could also be realized in cold atom setup \cite{Fujimoto} and the local bosonic couplings can also be achieved in cold atom system with tunable interacting bosonic environments. We assume the bosonic environments are also described by conformal invariant theory and take the special case of $AdS_5$ space Holographic theory. 
    
\subsection{the environmental influence function of helical Luttinger liquids and its CFT generalizations}    
           The typical environments composed of 1D electrons are either Fermi or Luttinger liquids. For the helical Luttinger liquids wires case the Majorana-dressed symmetric Green function with election chemical potential $\mu$ takes the following form (for more details, see Appendix \ref{A-1}.)
\be\label{symG-f}
\overline{G}_{a,sym}(\omega)= c_1(\kappa) |\omega -\mu|^{2\kappa-1}\; e^{-{\omega^2 \over \Gamma_0^2}}\;,\quad (\kappa\ge 1/2)\;,
\ee 
   
where $\kappa\equiv \left(K+\frac{1}{K}\right)/4$ with $K$ denoting the Luttinger parameter is related to conformal dimension of the operator $\mathcal{O}_a$ and is used to characterize the Luttinger liquid, e.g., $\kappa=1/2$ for Fermi liquid. The
rough estimate on the Luttinger parameter $K$ is given by $K^2\sim(1 + \frac{U}{2\epsilon_F})$, where $\epsilon_F$ is the Fermi energy
and $U\sim \frac{e^2}{\epsilon a_0}$ ($\epsilon$ being the dielectric constant and $a_0$ being the lattice length) is the characteristic Coulomb energy of the wire\cite{Kane}. Thus the value of $\kappa$ can be tuned by changing the effective repulsive/attractive short range interactions in the wire. $c_1(\kappa)$ is some function of $\kappa$ and its exact form is given in Appendix \ref{A-1}. In \eq{symG-f} we have introduced a windowed function $ e^{-\omega^2 / \Gamma_0^2}$ to cutoff the high frequency modes but the main results we mention below does not depend on this choice of cutoff.
 
   Similarly, for the bosonic operators $\mathcal{O}_{ab}$'s of conformal dimension $\Delta$, the associated Majorana-dressed symmetric Green function is (for more details, see Appendix \ref{A-2})
\be\label{symG-b}
\overline{G}_{ab,sym}(\omega)=c_2(\Delta)  |\omega|^{2\Delta-4} \; e^{-{\omega^2 \over \Gamma_0^2}}\;,\quad (\Delta \ge 2)\;, 
\ee
where $c_2(\Delta)$ is some analytic function of $\Delta$ and its explicit form is given in Appendix \ref{A-2}.

  With \eq{symG-f} and \eq{symG-b} we carry out the double time-integral, and the results are
\be
\int^t d\tau \int^t d\tau' \;\overline{G}_{a,sym}(\tau-\tau')= {c_1(\kappa) \over \Gamma(2\kappa)} I_{2\kappa-1}(t;\mu,\Gamma_0)
\ee
for $\mathcal{O}_a$, and
\be
\int^t d\tau \int^t d\tau' \;\overline{G}_{ab,sym}(\tau-\tau')=  c_2(\Delta)  I_{2\Delta-4}(t;0,\Gamma_0)
\ee
for $\mathcal{O}_{ab}$, with
\be\label{infoq}
I_{Q}(t;\mu,\Gamma_0):=\int_0^{\infty} (|\omega+\mu|^Q+|\omega-\mu|^Q) \; {2-2\cos \omega t  \over \omega^2}\; e^{-{\omega^2 \over \Gamma_0^2}}\; d\omega\;.
\ee

 For $\mu=0$ case, the integral in \eq{infoq} is worked out as (for $Q\ge 0$):
\beq \label{Iint}  
&&I_Q(t;0,\Gamma_0)=4\int_0^{\infty}|\omega|^Q\frac{1-\cos\omega t}{\omega^2}e^{-\frac{\omega^2}{\Gamma_0^2}} \\\nn
&&=\begin{cases}2\Gamma_0^{Q-1}\Gamma\left(\frac{Q-1}{2}\right)\left(1-\; _1F_1\left(\frac{Q-1}{2};\frac{1}{2};-\frac{t^2\Gamma_0^2}{4}\right)\right)\\\nn
\quad \text{for} \quad Q\ge 0  \quad \text{but} \quad Q\ne 1; \\\nn
\frac{1}{2}t^2 \Gamma_0^2 \; _2F_2\left(\{1,1\};\{\frac{3}{2},2\};-\frac{t^2\Gamma_0^2}{4}\right)\\\nn  \quad \text{for} \quad Q=1.\end{cases}
\eeq   
Here  $_pF_q(\{a_1,..,a_p\};\{b_1,..b_q\};z)=\sum_{k=0}^{\infty}\frac{(a_1)_k..(a_p)_k}{(b_1)_k..(b_q)_k}\frac{z^k}{k!}$ is the generalized hypergeometric function, and $\Gamma(z)$ is the Gamma function. 

    First, we see that the behavior of \eq{Iint} at $t\approx 0$ is universal, i.e., $I_Q \propto t^2$ independent of $Q$. Using this, it is straightforward to show the following universal Tsunami behavior of entanglement growth as discussed in \cite{Ho:2013rra,Liu:2013iza}:
\be
S_2 = -\ln \Tr \rho^2_r \propto t^2 \qquad \mbox{for}\quad t\approx 0\;. 
\ee
Note that $S_2$ is the 2nd order Renyi entropy, which characterizes the quantum entanglement between the probe and environment.  The above $t^2$ growth is more rapid than the expected $t$ growth by causality argument, and thus called Tsunami behavior. However, this Tsunami behavior will soon be taken over by $t$ growth once the spreading scale is above the size of the probe. 
    
  \begin{figure} 
\includegraphics[width=0.85\columnwidth]{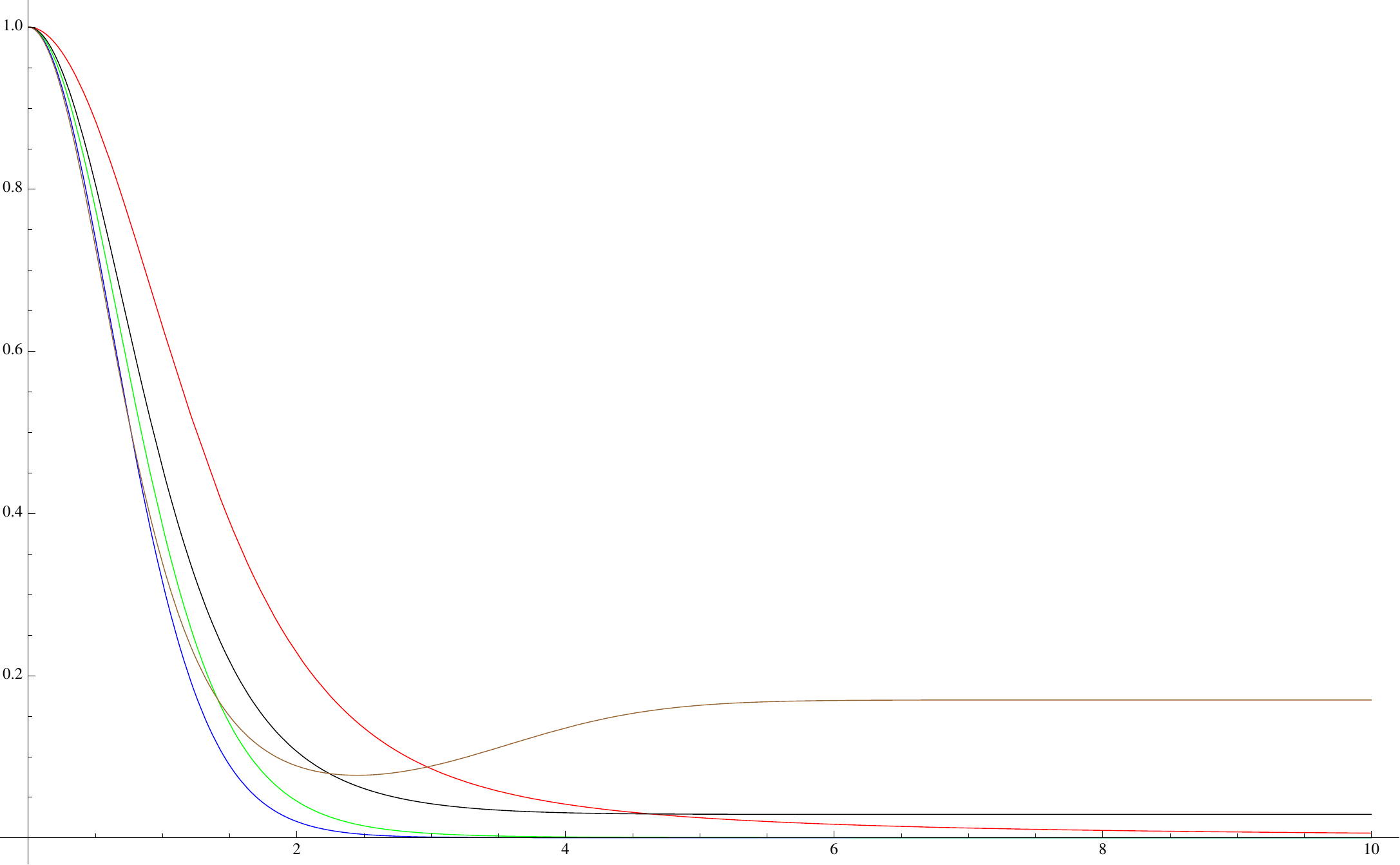}
\caption{$e^{-I_Q(t;\mu=0,\Gamma_0=1)}$ v.s. $t$ for $Q=0.5$(blue), $0.9$(green), $1$(red), $2$ (black) and $4$ (brown). This factor controls the time dependence of the influence functional. We can see that  there is a critical value at $Q=1$ beyond which this factor will have a pattern  of drop-dip-flat and will not decay to zero. }
\label{alphaQ}
\end{figure}

    On the other hand, the late time behavior of \eq{Iint} has an interesting turning point at $Q=1$, i.e., $\kappa=1$ or $\Delta=2.5$. For $Q\le 1$ (Ohmic and sub-Ohmic environments) the magnitude of the hypergeometric function in \eq{Iint} increases monotonically with increasing time $t$. However, for $Q>1$ (super-Ohmic environments) it decreases to zero at large time so that $I_Q(t;0,\Gamma_0)\rightarrow 2\Gamma_0^{Q-1}\Gamma\left(\frac{Q-1}{2}\right)$ as $t\rightarrow\infty$.  See Fig. \ref{alphaQ} for $e^{-I(t;\mu=0,\Gamma_0)}$ for this change of behavior as varying $Q$. This qualitative change of the time dependence has two implications:
    
 \begin{itemize}
 \item There should be a metal-to-insulator like quantum phase transition, similar to the case of electric transport \cite{SungPo1}, by tuning $Q$ for the probe state due to back reaction of integrating out the environment. For $Q\le 1$, the number of the effective carriers quantified by \eq{Iint} is sufficient so that the probe state is de-localized and dissipative. On the contrary, for $Q>1$, the effective carriers is insufficient so that the state is localized and non-dissipative.

 \item As the probe carries both energy and quantum information, the de-localization of the probe state also implies the leakage of the quantum information into the environment. This finally leads to complete decoherence. Otherwise, the quantum information is confined around the localized probe state and only partially leaks into the environment. In In such a case, one may further purify the probe state to recover and maintain the quantum information carried by the probe. Though our interpretation of the incomplete decoherence for our setup  is quite different from the usual discussion of non-Markovian dynamics which is mainly due to due to the gap-like structures of the environmental spectral densities \cite{ZhangPRL}, we do see the similar information backflow for $Q>1$ cases (as seen from the drop-dip-flat patterns of the purity shown below) characterizing the  non-Markovian dynamics.

\end{itemize}

      The above picture of quantum phase transition might be understood by the Renormalization Group (RG) argument of the coupling constant $B_{M}$. Note that the fermionic environment we choose in this paper is the helical Luttinger liquids, which can be realized as the interacting edge state of two dimensional Topological insulator \cite{Zhang,SungPo1}. For this kind of fermionic environment, the scaling dimension from zeroth order renormalization group analysis \cite{SungPo1,Gogolin} in interaction $\hat{V}$ is $\kappa$. This renders the renormalized coupling $B_a$ in linear response regime as
\be\label{rg eq}
\frac{dB_a}{d\ln(\Gamma_0)}=\left(1-\kappa\right)B_{a} \quad,
\ee
with $\Gamma_0\simeq\hbar v/a_0$ as the UV cutoff for the liner spectrum of the edge state. For the electric transport discussed in \cite{SungPo1} $\kappa=1$ indicates
the critical interaction strength for metallic to insulating behavior. That is, following \eq{rg eq} we see that for $1\ge\kappa\ge 1/2$ the coupling $B_a$ increases as cutoff decreases. While for $\kappa>1$ the coupling $B_a$ decreases with decreasing cutoff, indicating insulating behavior for charge transport.

\subsection{Pointer States, Purity and Concurrence}
       
     Though the time dependence of \eq{Iint} provide a very suggestive picture for the change of the decoherence pattern by tuning $Q$, one should still study the detailed form of the final reduced density matrix, from which we derive some quantities to characterize the decoherence patterns.
     
       There are many ways to characterize the quantum decoherence. If the probe state is in the pointer-state basis \cite{Environment}, then we can characterize the decoherence by observing the time evolution of the off-diagonal elements of the reduced density matrix. If all the off-diagonal elements vanish in the long run, then the final reduced density matrix is in the pointer state. In this case, the probe state decoheres completely. Among the pointer states, a special one is the Gibbs state with all the diagonal elements equal. In such a case, we can furthermore claim that the probe state has been thermalized (at zero temperature).

       However, if the final state does not reduce to a pointer state, then we shall find some quantity to characterize the quantumness of the probe state. In this work we choose the purity 
\be \label{purityP}     
\mathcal{P}^{b,f}:=\Tr \rho^2, 
 \ee    
where the superscripts $b,f$ refer to either bosonic or fermionic environments. The purity is related to the 2nd order Renyi entropy by $S_2:=-\ln \mathcal{P}^{b,f}$.

  Note that, without knowing if all the off-diagonal elements of the reduced density matrix diminish or not, one cannot tell by the purity alone if the qubits decohere completely unless for the Gibbs state. The purity of the N-qubit state reaches its minimal value ${1\over N}$ for the Gibbs state. Thus, a thermalized state is definitely a classical state.

   For multi-qubit cases, one can also characterize the decoherence by the time evolution of the quantum entanglement among the qubits. For the two-qubit case this is faithfully characterized by the concurrence \cite{Concurrence}:
\be
\mathcal{C}(\rho) := \max(0, \lambda_1-\lambda_2-\lambda_3-\lambda_4)
\ee
 where $\lambda_1, \cdots, \lambda_4$  are the square roots of the eigenvalues, in the decreasing order, of $\rho\tilde{\rho}$ with $\tilde{\rho}:= (\sigma_y \otimes \sigma_y) \rho*  (\sigma_y \otimes \sigma_y)$ for $\sigma_{y}$ to be the Pauli matrix.  If the final probe state is a pointer state, i.e., completely decohered, then $\mathcal{C}=0$ at large time as the quantum entanglement is an essential property only for quantum state. But the inverse is not true as the quantum state may not be entangled.

\subsection{Cases of single topological qubit}\label{singleq}

   In the first case, we consider only two Majorana modes $\gamma_1$ and $\gamma_2$ living on two ends of a quantum wire, which couple locally to the surrounding electrons via the channels $\mathcal{O}_E$ appearing in the interaction \eq{majV}.  Before turning on the interaction \eq{majV}, these two Majorana modes form a topological (non-local) qubit with state $|0\rangle$ and $|1\rangle$ connected by
\be
{1\over 2}(\gamma_1 - i \gamma_2)|0\rangle= |1\rangle\;, \qquad {1\over 2}(\gamma_1 + i \gamma_2)|1\rangle= |0\rangle\;.
\ee
 We can choose the following representation for $\gamma_{1,2}$, viz., 
\be\label{parityop}
\gamma_1=\sigma_1\;, \qquad \gamma_2 = \sigma_2\;, \qquad i \gamma_1\gamma_2= \sigma_3
\ee
where $\sigma_a$'s are the Pauli matrices so that they satisfy $\gamma_a^{\dagger}=\gamma_a$ and $\{\gamma_a,\gamma_b\}=2\delta_{ab}$. Note that $i\gamma_1\gamma_2$ defines a parity operator for the states of the topological qubit, i.e., 
\be
i\gamma_1\gamma_2|0\rangle=|0\rangle\;, \qquad i\gamma_1\gamma_2|1\rangle=-|1\rangle\;,
\ee
Obviously, the interactions $\gamma_1 \mathcal{O}_1$ and $\gamma_2 \mathcal{O}_2$ flip the parity of the topological qubit but $\gamma_1 \gamma_2 \mathcal{O}_{12}$ does not.

In this representation, the generic initial state of the probe can be casted into a hermitian matrix with positive eigenvalues
\beq \label{inirho-1}
\rho_{\mathcal{P}}(t=0)=\left(\begin{array}{cc} a_{00} & a_{01} \\ a_{01} & a_{11} \end{array}\right)  
\eeq
with $a_{00}+a_{11}=1$.

 Here we consider either (i) fermionic environmental channel with the evolution operator $U(t)=\textrm{T} e^{-i \gamma_1 {\bf O}_1(t)} e^{-i \gamma_2 {\bf O}_2(t)}$ or (ii) bosonic environmental channel with $U=e^{-i \gamma_1\gamma_2  {\bf O}_{12}(t)}$. Using \eq{rhofinal-s} the explicit form of the reduced density matrix at time $t$ is 
\begin{widetext} 
\beq \label{frho-1}
\rho^f_r (t)={1\over N^f (t)}\left(\begin{array}{cc} a_{00} (C_1 C_2+S_1 S_2)-a_{11} (C_2 S_1+C_1 S_2) & a_{01} (C_1 C_2-S_1 S_2)-a_{10} (C_2 S_1-C_1 S_2) \\ a_{10} (C_1 C_2-S_1 S_2)-a_{01} (C_2 S_1-C_1 S_2) & a_{11} (C_1 C_2+S_1 S_2)-a_{00} (C_2 S_1+C_1 S_2) \end{array}\right)  \nn \\
\eeq
\end{widetext}
for fermionic channel, and 
\beq \label{brho-1}
\rho^b_r (t)={1\over N^b (t)}\left(\begin{array}{cc} a_{00} (C_{12}-S_{12}) & a_{01} (C_{12}+S_{12}) \\ a_{10} (C_{12}+S_{12}) & a_{11} (C_{12}-S_{12}) \end{array}\right) 
\eeq
for bosonic channel. In the above, the normalization factor $N^{f,b}(t)=\Tr \; \rho^{f,b}_r(t)$, and 
\bea
C_M &\equiv& \langle \tilde{\textrm{T}} \cosh{\bf O}_M(t) \textrm{T} \cosh{\bf O}_M(t) \rangle_{\mathcal{E}}\;,\\
S_M &\equiv& \langle \tilde{\textrm{T}} \sinh{\bf O}_M(t) \textrm{T} \sinh{\bf O}_M(t) \rangle_{\mathcal{E}}\eea
for either $M=a$ or $ab$. Their relations with the Majorana-dressed symmetric Green functions are given in
\eq{ccG} and \eq{ssG}. 

  Note that there is a key difference between \eq{frho-1} and \eq{brho-1}. In \eq{frho-1} the states with different parities, i.e., $|0\rangle\langle 0|$ and $|1\rangle\langle 1|$, or  $|0\rangle\langle 1|$ and $|1\rangle\langle 0|$, mix as time evolves, but this does not happen in \eq{brho-1}. This is because the reason as stated below \eq{parityop}, i.e., the fermionic environments flip the parity of the topological qubit but the bosonic ones do not.

  We further simplify the above reduced density matrices by assuming all the Majorana modes couple to the same environments with the same coupling strengths. Under this condition we can omit the sub-index $M$ of $\overline{G}_{M,sym}$ and $B_M$ as they all are the same. The reduced density matrix is then simplified to
\beq \label{frho-2}
\rho^f_r (t)={1\over 2}\left(\begin{array}{cc}1+(2 a_{00}-1)\alpha^2(t) & 2 a_{01}\alpha(t) \\  2 a_{10}\alpha(t) & 1+(2 a_{11}-1)\alpha^2(t) \end{array}\right)  \nn \\
\eeq
and 
\beq \label{brho-2}
\rho^b_r (t)=\left(\begin{array}{cc} a_{00} & a_{01}\alpha(t) \\ a_{10} \alpha(t) & a_{11}  \end{array}\right) 
\eeq
where the influence functional 
 \be\label{influence-alpha}
 \alpha(t)= e^{2 B^2 \int^t d\tau \int^t d\tau' \overline{G}_{sym}(\tau-\tau') }=e^{-2B^2|\alpha_{1,2}| I_Q(t;\mu=0,\Gamma_0)}\;.
 \ee
where $\alpha_{1,2}$'s are the time-independent overall coefficients in front of the Majorana-dressed symmetric Green functions given in \eq{A04} and \eq{B02-2}, respectively.  Note that these coefficients are negative.

   The typical behavior of $\alpha(t)$ for different $Q$ can be inferred from in Fig. \ref{alphaQ}. The factor $2B^2|\alpha_{1,2}|$ in \eq{influence-alpha}  will only affect the behaviors in  Fig. \ref{alphaQ} quantitatively but not qualitatively.  Thus, there is a critical point at $Q=1$ for $\alpha(t)$. For $Q\le 1$ (Ohmic and sub-Ohmic), $\alpha(t)$ vanishes at large $t$ so that the state of the topological qubit reduces to a pointer state both in the fermionic and bosonic environments. This implies that the topological qubit decohere completely. However, there is a key difference for fermionic and bosonic cases: the pointer state in the (parity-flipping) fermionic environments is the Gibbs state, but not the case in the (parity-conserving)  bosonic ones. The reduction to the Gibbs state means  thermalization (in the sense of micro-canonical ensemble). The thermalization is due to the mixing of the parity odd and even states causing by the parity-violating interactions. In contrast, there is no mixing in the bosonic case so that the diagonal elements of \eq{brho-2} remain constant without going into $1/2$ as the off-diagonal elements diminish. We can then conclude that in the sub-Ohmic environments, though the parity conservation cannot prevent a topological qubit from complete decoherence, it can prevent it from complete thermalization \cite{parity-issue}.  Note that the term ``thermalization" here is used in the context of micro-canonical ensemble for our zero temperature setup or in the context of closed system (probe plus environment) thermalization. For considering the topological qubits in contact with the  thermal environment,  a new observable characterizing the topological nature of the Majorana modes should be constructed \cite{Uhlmann}.

   For both types of environments at $Q>1$ (super-Ohmic), $\alpha(t)$ reduces to a nonzero constant at late time so that the off-diagonal elements of the reduced density matrices do not vanish. This implies that the probe state cannot completely decohere, and the incomplete decoherence can be quantified by the purity \eq{purityP}.  
   
   In the fermionic cases, the parameter $\kappa$ characterizes the interaction strengths of the Luttinger/Fermi liquid. The larger $\kappa$ is, the stronger interaction/correlation is shown in the Luttinger liquid nanowire. Our above results suggest that the strongly correlated environments help to preserve the quantum information of the probe even the probe-environment coupling is weak. This could be understood as follows: the weakly coupled probe-environment contact spot becomes a peculiar point in contrast to their strongly coupled neighbors so that the quantum information stay around this particular place.

In section \ref{non-uniform} we consider the cases of non-uniform environments.  We still keep the coupling strengths $B_M=B$ uniform to focus on the effect of varying the environmental spectral functions $\overline{G}_{M,sym}$ with $M$. Tuning $B_M=B$ is achievable in the experiments by controlling different tunneling junctions' gate voltages.

\subsection{Non-uniform environments}\label{non-uniform}     

   As the topological qubit is non-local, it is interesting to consider a peculiar case which cannot happen for the usual local qubit, that is, the non-uniform environment. In this case, the two Majorana modes can couple to the environment of different $\kappa$'s, i.e., $\kappa_1$ and $\kappa_2$. Given the initial state \eq{inirho-1}, the reduced density matrix $\rho^f_r(t)$ with two different $\kappa$'s takes the form 
\beq\label{f-nonu-rho}
\left(\begin{array}{cc} \frac{1}{2}+\alpha_1(t)\alpha_2(t) (a_{00}-\frac{1}{2}) &  \alpha_2(t) \Re a_{01} + i \alpha_1(t)\Im a_{01}  \\   \alpha_2(t)\Re a_{01}-i \alpha_1(t)  \Im a_{01} & \frac{1}{2}-(a_{00}-\frac{1}{2})\alpha_2(t) \alpha_1(t)\end{array}\right).
\eeq  
from which one can obtain the purity as follows:
\be\label{Pnon-u}
\mathcal{P}^f(t)=\frac{1}{2} + 2 [\Im a_{01}]^2 \alpha_1^2(t)(1-\alpha_2^2(t)) +2 [\Re a_{01}]^2\alpha_2^2(t)  
\ee
where 
\bea
\alpha_1(t)&=& e^{2 B^2 \int^t d\tau \int^t d\tau' \overline{G}_{1,sym}(\tau-\tau') }\;,\\
\alpha_2(t)&=& e^{2 B^2 \int^t d\tau \int^t d\tau' \overline{G}_{2,sym}(\tau-\tau') }
\eea
with $\overline{G}_{1,sym}$ and $\overline{G}_{2,sym}$ corresponding to the Majorana-dressed symmetric Green functions of $\kappa_1$ and $\kappa_2$, respectively. In arriving \eq{Pnon-u} we have used the fact that $\rho_{\mathcal{P}}(t=0)$ is hermitian so that $a_{10}=a^*_{01}$. As $\alpha_{1,2}^2(t)$ are always smaller than one so that the expression \eq{Pnon-u} implies $\mathcal{P}^f(t)\ge 1/2$.  Note that $\mathcal{P}^f=1/2$ corresponds to the Gibbs state.

    From \eq{f-nonu-rho}, we see that if both $\alpha_{1,2}$ have the same late time behaviors, then the decoherence patterns are qualitatively similar to the cases with uniform environments. Otherwise we find new decoherence patterns. More specifically, for uniform environments the final state purity depends both on $\Im a_{01}$ and $\Re a_{01}$. On the other hand, for the non-uniform environment with different late time behaviors, \eq{f-nonu-rho} shows the bias, that is, the the first channel prefers the initial state with non-zero $\Im a_{01}$ to be robust against complete decoherence, and the second channel prefers the one with non-vanishing $\Re a_{01}$.
It suggests that the  informations of real and imaginary part of off-diagonal reduced density matrix element $a_{01}$ are carried by different end points of Majorana modes. Thus, the dereference pattern not only depend on the environments but also on the initial states.  Note that this kind of non-uniformity of the environments is a peculiar feature of the topological qubit composed by Majorana modes.

   We would like to illustrate the implication of the above results more here. In the context of closed system, all the single qubit pure states are connected by unitary transformation and thus are equivalent. For local qubits these unitary transformation can be performed locally, and for the topological qubits non-local unitary operations are needed. However, once the qubits are put in contact with the environments, the unitarity get lost as time evolves and the naive equivalence of all the pure states no longer holds. The loss of unitarity implies that the democracy of pure states representing quantum information is broken, and some states are more robust against decoherence than the others.  From our above discussions, we see that the non-uniformity of the environments further enhance the inequality among the pure states. The state space of a single qubit is represented by Bloch sphere, and the breaking of state democracy implies the spontaneously breaking of the isometry of Bloch sphere. The non-uniformity of the environments breaks the isometry more badly. 
     
 As shown, the decoherence pattern of a single topological qubit can be read out directly from the time dependence of the reduced density matrix. This is not the case for the generic initial states of two topological qubits (formed by four Majorana fermions) due to the complexity of the analytic formula. We numerically plot the purity and concurrence to characterize the decoherence patterns for those cases in section \ref{2qubits}.
              
\subsection{Cases of two topological qubits}\label{2qubits}

  Now we consider the the cases of two topological qubits made of four Majorana modes. The Hilbert space of two qubits are spanned by $\{ |00\rangle, |01\rangle, |10\rangle, |11\rangle \}$, which are connected by
\bes\label{4m-2q}
\beq 
&& d_1^{\dagger} |00 \rangle \equiv \frac{1}{2} (\gamma_1 - i  \gamma_2 ) |0 0 \rangle =|10 \rangle\;, \\
&& d_2^{\dagger} |00 \rangle \equiv \frac{1}{2} (\gamma_3 - i  \gamma_4 ) |0 0 \rangle =|01 \rangle\;, \\
&& d_1^{\dagger}d_2^{\dagger} |00 \rangle  =|11 \rangle \;.
\eeq
\ees

  Generalizing the single topological qubit case, here we can define two commuting parity operators for the states of two topological qubits, i.e.,
\be\label{2parity}
i\gamma_1\gamma_2 |j k\rangle = (-1)^j |j k\rangle\;, \qquad  i\gamma_3 \gamma_4 |j k\rangle = (-1)^k |j k\rangle
\ee
with $j,k=0,1$.  Note that other two-gamma operators  such as $i\gamma_1\gamma_{3}$ are not the parity operators. From these facts that, we know the interactions $\gamma_1\gamma_2 \mathcal{O}_{12}$ and $\gamma_3\gamma_4 \mathcal{O}_{34}$ preserve the parity of the topological qubit but the other interactions do not.  Furthermore, if we consider also interaction involving four Majorana modes, i.e., $\gamma_1\gamma_2\gamma_3\gamma_4 \mathcal{O}_{1234}$, we may also introduce another parity operator $\Gamma^{(4)}:=-\gamma_1\gamma_2\gamma_3\gamma_4$ which also commutes with the previous two parity operators, i.e.,
\be\label{gamma4}
\Gamma^{(4)}|j k\rangle = (-1)^{j+k}|j k\rangle\;.
\ee
Thus, $|00\rangle$ and $|11\rangle$ are $\Gamma^{(4)}$-parity even states and $|01\rangle$ and $|10\rangle$ are odd states. This echoes the choice of using $|00\rangle$ and $|11\rangle$ as a single topological qubit\cite{Chamon,Budich,Loss} if we deemed the effective p-wave superconductor in the Kitaev chain as a kind of external bosonic environment for the Majorana end points.

   As the Majorana fermions obey the Clifford algebra: $\{\gamma_i,\gamma_j\}=2\delta_{ij}$, we can then choose the following representation
\beq\nn 
&&\gamma_1=\sigma_1\otimes \sigma_0 \;, \qquad \gamma_2 = \sigma_2\otimes \sigma_0 \;,\\ &&\gamma_3 = \sigma_3 \otimes \sigma_1  \;, \qquad \gamma_4 = \sigma_3 \otimes \sigma_2.
\eeq
  
    The generic initial density matrix for the probe is given by a $4\times 4$ hermitian matrix with positive eigenvalues.  For general initial state, the explicit form of $\rho_r(t)$ following from \eq{rhofinal-s} is very tedious. Since we are considering the quantum decoherence, we take the initial state as a pure state with the following form 
\be\label{sgeneric}
|(e_1,e_2, e_3, e_4) \rangle =e_1 |00\rangle+e_2 |01\rangle+e_3 |10\rangle+e_4 |11\rangle
\ee
with $|e_1|^2+|e_2|^2+|e_3|^2+|e_4|^2=1$.

     In the following we assume the fermionic channel takes the form $\gamma_i\mathcal{O}_i$ and the bosonic channel takes the form $\gamma_1\gamma_2\mathcal{O}_{12}$ and  $\gamma_3\gamma_4\mathcal{O}_{34}$ which are parity-preserving.  All the numerical plots shown below are for these interactions.
     
       On the other hand, the effect of the bosonic  parity-violating couplings such as $\gamma_1\gamma_3\mathcal{O}_{13}$ and $\gamma_2\gamma_4\mathcal{O}_{24}$ is briefly discussed here. We wonder if these interactions will mix the different parity sectors and lead to thermalization in the (uniform) sub-Ohmic environments as for the fermionic cases. The answer is no as seen from the final state density matrix for the given initial state \eq{sgeneric}, i.e., 
\beq \label{reduced-D-s2}   
\rho^b_r(\infty)= \left(\begin{array}{cccc}\frac{|e_1|^2+|e_4|^2}{2} & 0 & 0 & 0 \\0 & \frac{|e_2|^2+|e_3|^2}{2} & 0 & 0 \\0 & 0 & \frac{|e_2|^2+|e_3|^2}{2} & 0 \\ 0 & 0 & 0 & \frac{|e_1|^2+|e_4|^2}{2}\end{array}\right)\;.
\eeq 
\eq{reduced-D-s2} shows that parity-violating bosonic couplings mix even parity states with other even parity states but not odd ones (similarly for odd parity states can only mix with other odd parity states) \cite{parity-issue}.

\begin{figure}
\includegraphics[width=0.85\columnwidth]{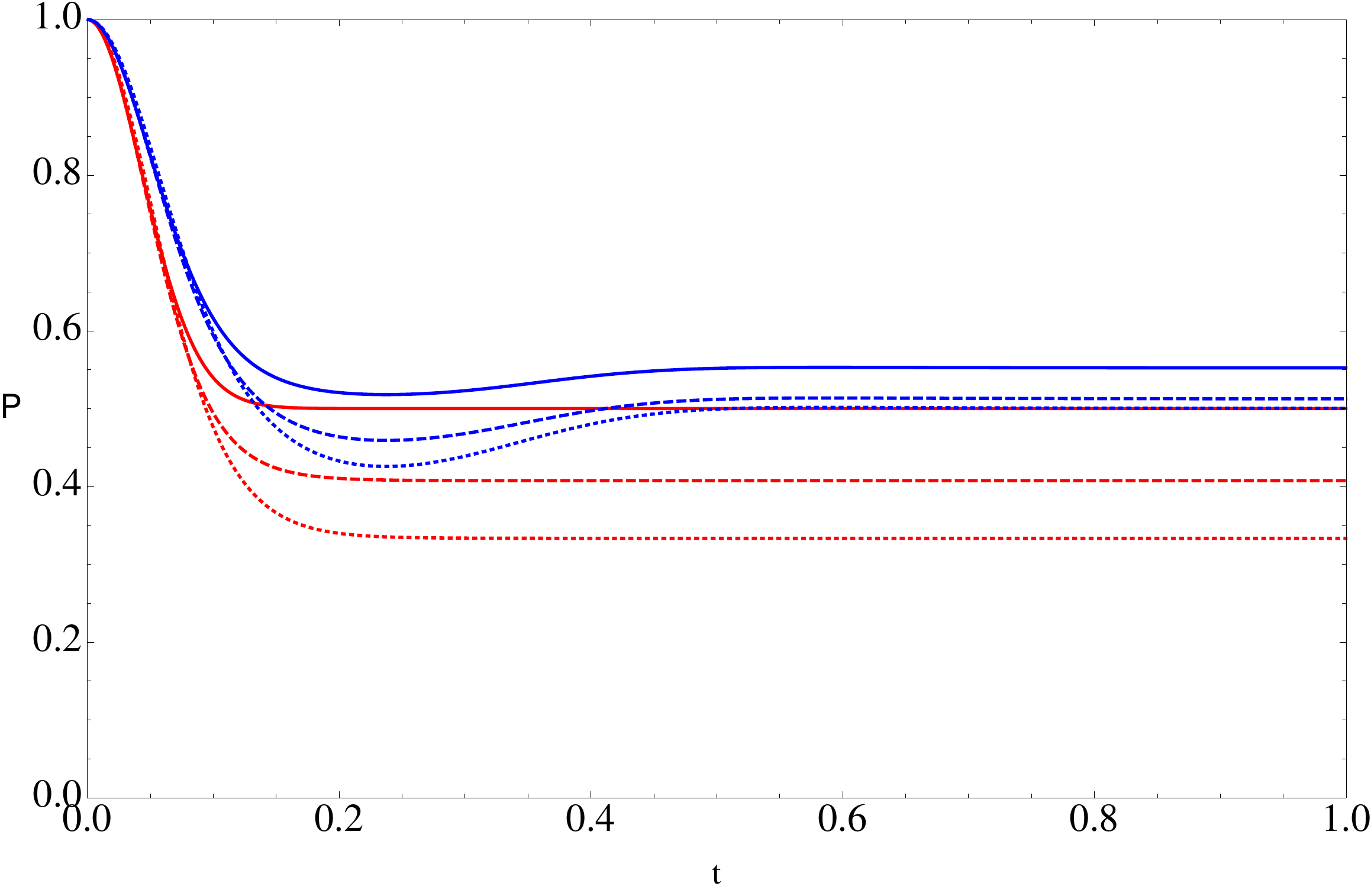}
\caption{Purity vs $t$ for $\Delta= 2.3$ (red) and $\Delta=4.1$ (blue) with initial states $|(e_1,e_2, e_3, e_4) \rangle=|(1,0,0,1)\rangle$ (solid), $|(2,1,0,2)\rangle$ (dashed), $|(1,1,0,1)\rangle$ (dotted). }
\label{PbEs}
\end{figure}

\begin{figure}
\includegraphics[width=0.85\columnwidth]{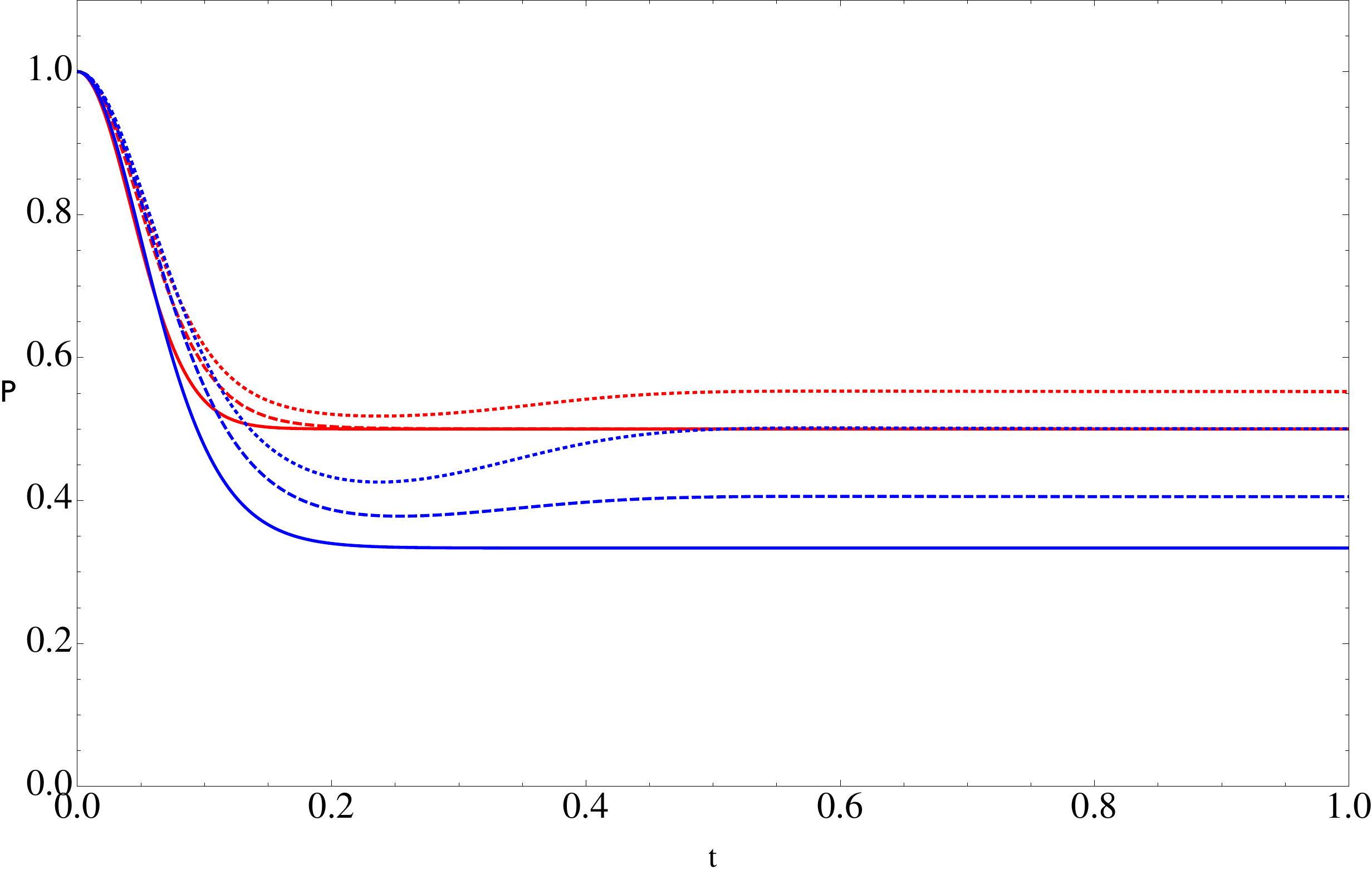}
\caption{Purity vs $t$ for non-uniform bosonic environments with initial states $|(e_1,e_2, e_3, e_4) \rangle=|(1,0,0,1)\rangle$ (red), $|(2,1,0,2)\rangle$ (blue).  The environments are specified by the $\Delta$ vector: $(\Delta_{12},\Delta_{34})= (2.3,2.3)$ (solid), $(2.3,4.1)$ (dashed) and $(4.1,4.1)$ (dotted). }
\label{PbNs}
\end{figure}

   Now we will start our numerical case studies. For illustration, we first consider a special case with the initial state of the topological qubits being $|(e_1,0, 0, e_4) \rangle$, which corresponds to the initial state density matrix
\be\label{2qbell}
\rho_{\mathcal{P}}(t=0)=\left(\begin{array}{cccc}|e_1|^2 & 0 & 0 & e_1e_4^* \\0 & 0 & 0 & 0 \\0 & 0 & 0 & 0 \\e_1^* e_4 & 0 & 0 & |e_4|^2\end{array}\right)\;.
\ee
Furthermore, we assume the fermionic channel takes the form $\gamma_i\mathcal{O}_i$ and the bosonic channel takes the form $\gamma_1\gamma_2\mathcal{O}_{12}$ and  $\gamma_3\gamma_4\mathcal{O}_{34}$.  For such cases, the reduced density matrix are given in \eq{D03} for the fermionic channel and in \eq{D04} for the bosonic channel. If we consider the uniform environment as in the single qubit case, then we can further simplify the reduced density matrix. The results are as follows:
\beq \label{reduced-D-f} 
\rho^f_r(t)=\frac{1}{4}\left(\begin{array}{cccc}A_{11} & 0 & 0 & A_{14} \\0 & A_{22}  & 0 & 0 \\0 & 0 & A_{33}  & 0 \\ A_{41} & 0 & 0 & A_{44} \end{array}\right)
\eeq
with $A_{11}=1+\alpha(t)^4+2(2 |e_1|^2-1)\alpha(t)^2$, $A_{22}=A_{33}=1-\alpha(t)^2$, $A_{44}=1+\alpha(t)^4+2(2 |e_4|^2-1)\alpha(t)^2$ and $A_{14}=A_{41}^{\ast}=4 e_1 e_4^* \alpha(t)^2$
and
\beq \label{reduced-D-s}
\rho^b_r(t)= \left(\begin{array}{cccc}|e_1|^2 & 0 & 0 & e_1e_4^{\ast}\alpha(t)^2 \\0 & 0 & 0 & 0 \\0 & 0 & 0 & 0 \\ e_1^{\ast}e_4\alpha(t)^2 & 0 & 0 & |e_4|^2 \end{array}\right)\;.
\eeq  
Similar to the single qubit case discussed in section \ref{singleq}, for $Q>1$ the topological qubits cannot decohere completely for both fermionic and bosonic environments. For $Q\le 1$  $\rho^f_r(t)$ reduces to the Gibbs state and $\rho^b_r(t)$ reduces to a special pointer state at large time, indicating the topological qubits decohere completely but only thermalize in the fermionic environments. Thus with bosonic (parity-preserving and parity-violating) couplings the topological qubits decohere but in general does not reach thermalized states in the sub-Ohmic environments as shown later in Fig.\ref{PbEs} and Fig.\ref{PbNs}, in which the purity never reaches $1/4$ even for sub-Ohmic lines.

\begin{figure}
\includegraphics[width=0.95\columnwidth]{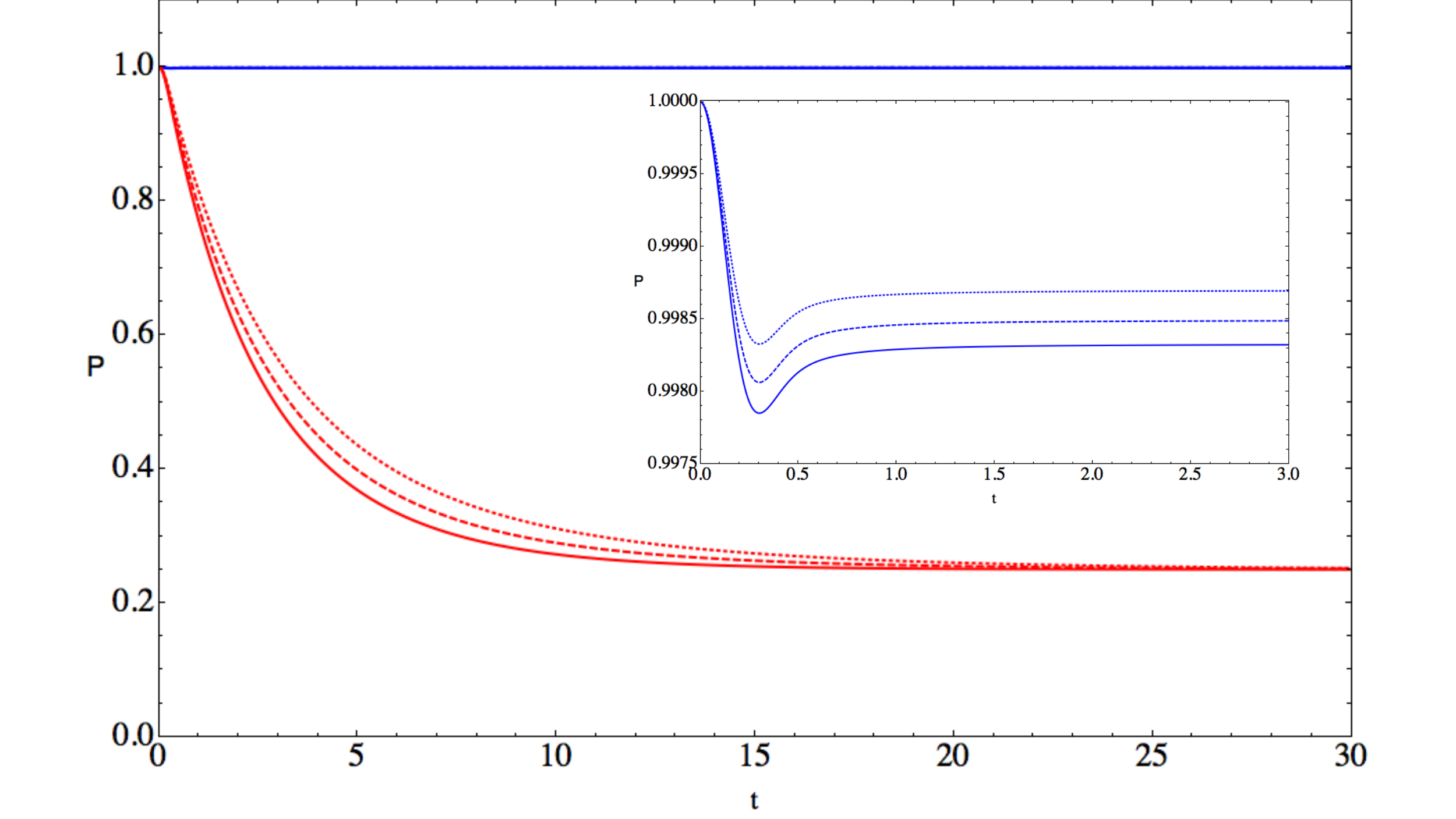}
\caption{Purity vs $t$ for $\kappa= 0.5$ (red) and $\kappa=2$ (blue) with initial states $|(e_1,e_2, e_3, e_4) \rangle=|(1,0,0,1)\rangle$ (solid), $|(2,1,0,2)\rangle$ (dashed), $|(1,1,0,1)\rangle$ (dotted). The inset is to magnify the early time region of $\kappa=2$ cases. }
\label{PfEs} 
\end{figure}  

\begin{figure}
\includegraphics[width=0.95\columnwidth]{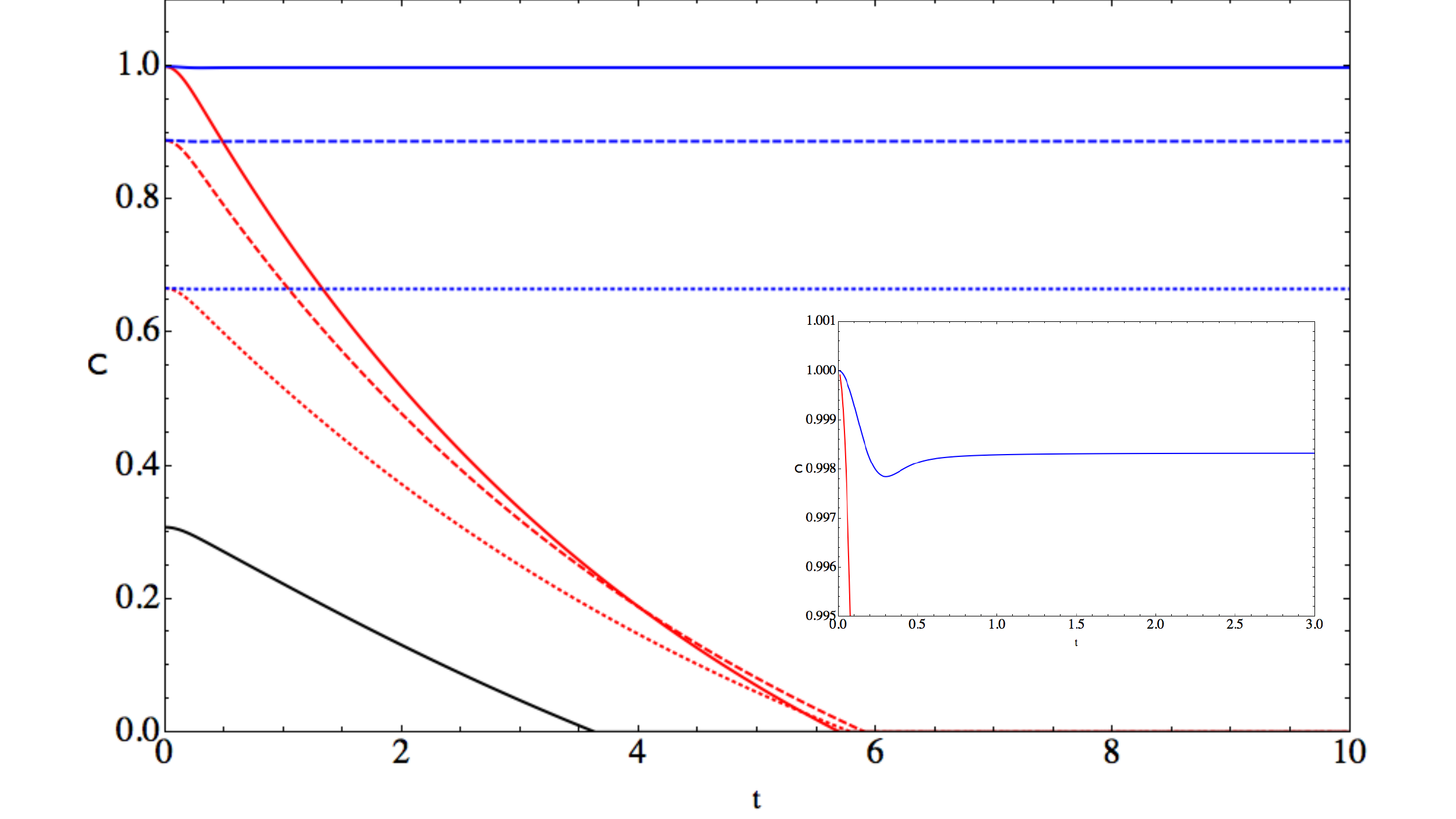}
\caption{Concurrence vs $t$ for the states and environments specified in Fig. \ref{PfEs}. The inset shows the solid lines enlarged at short time.  Here we add a black solid line representing the concurrence pattern of the initial state  $|(e_1,e_2, e_3, e_4) \rangle=|(2,2,1,2) \rangle$ in the $\kappa=0.5$  environment to show its concurrence does not diminish with the other red lines at the same time.}
\label{CfEs}
\end{figure}
 
\begin{figure}
\includegraphics[width=0.85\columnwidth]{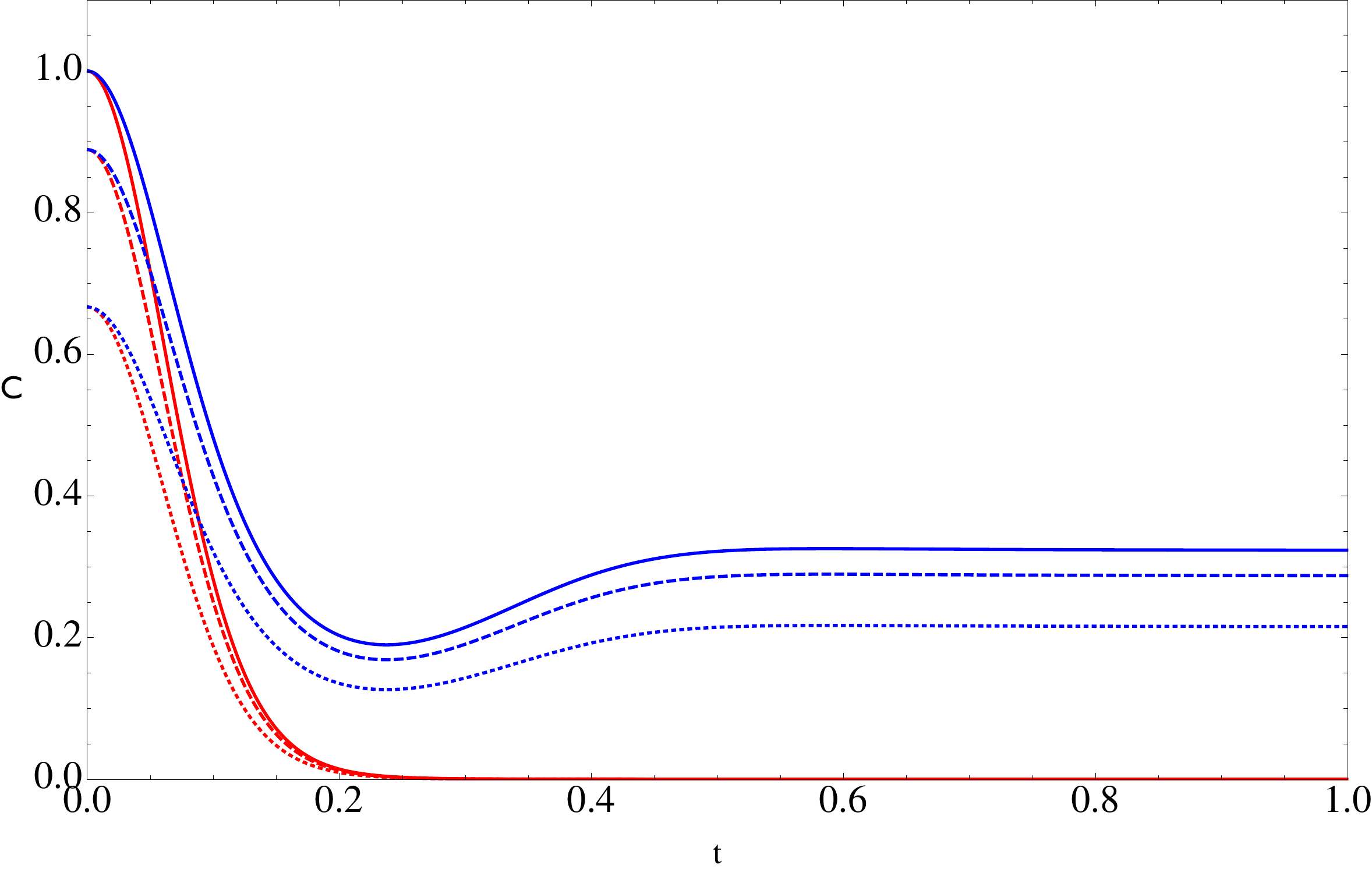}
\caption{Concurrence vs $t$ for the states and environments specified in Fig. \ref{PbEs}.}
\label{CbEs}
\end{figure}

\begin{figure}
\includegraphics[width=0.85\columnwidth]{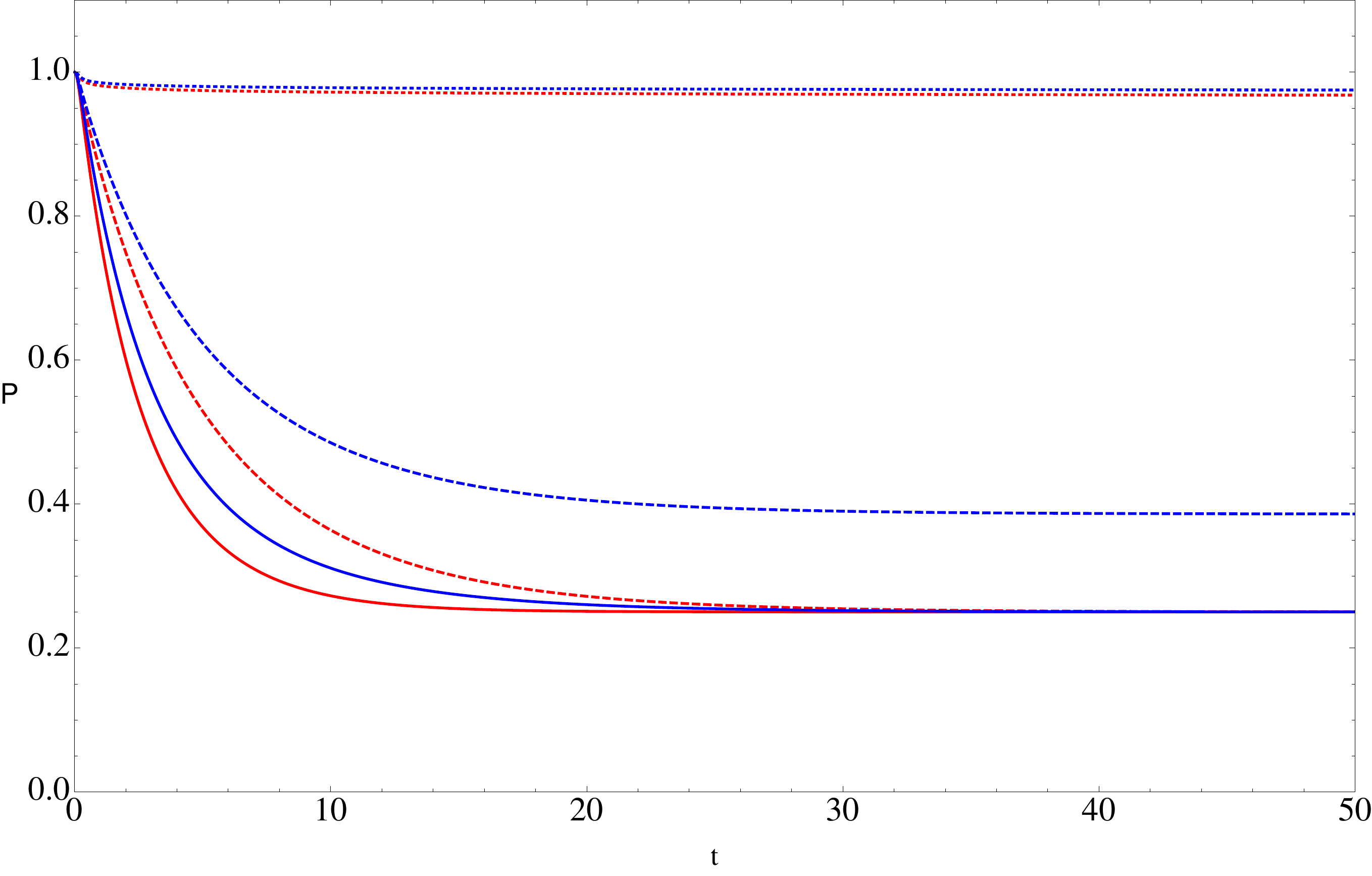}
\caption{Purity vs $t$ for non-uniform fermionic environments with initial states $|(e_1,e_2, e_3, e_4) \rangle=|(1,0,0,1)\rangle$ (red), $|(2,1,0,2)\rangle$ (blue).  The environments are specified by the $\kappa$ vector: $(\kappa_1,\kappa_2, \kappa_3, \kappa_4)= (0.5,0.5,0.5,0.5)$ (solid), $(0.5,0.5,1.2,1.2)$ (dashed) and $(1.1,1.1,2,2)$ (dotted). }
\label{PfNs}
\end{figure}

\begin{figure}
\includegraphics[width=0.85\columnwidth]{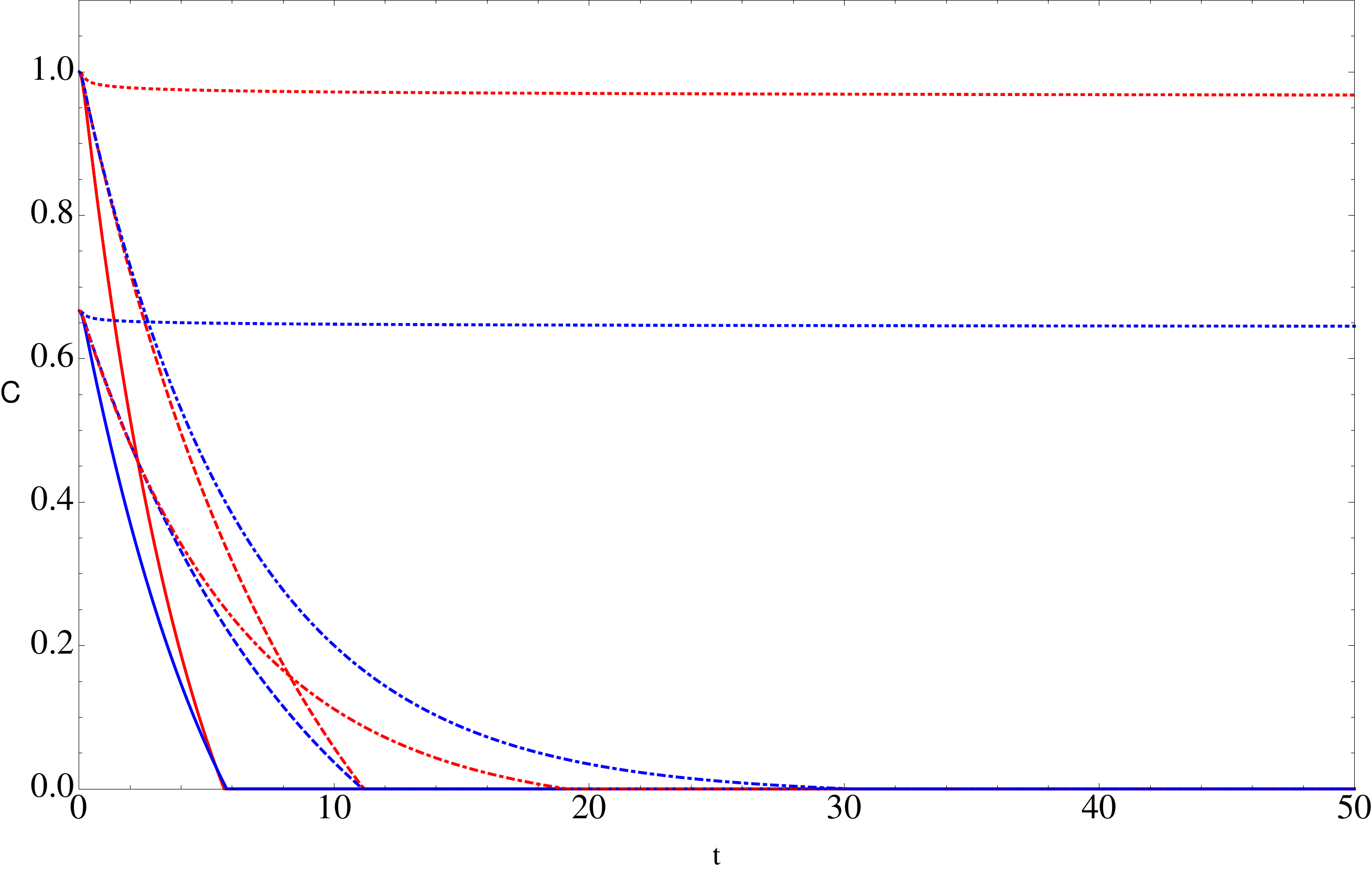}
\caption{Concurrence vs $t$ for the states and environments specified in Fig. \ref{PfNs}. Here we add the blue and red dot-dashed lines or the initial states $|(e_1,e_2, e_3, e_4) \rangle=|(1,0,0,1)\rangle$ (red), $|(2,1,0,2)\rangle$ (blue) in the environment with $(\kappa_1,\kappa_2, \kappa_3, \kappa_4)= (0.5,1.2,1.2,0.5)$. }
\label{CfNs}
\end{figure}

\begin{figure}
\includegraphics[width=0.85\columnwidth]{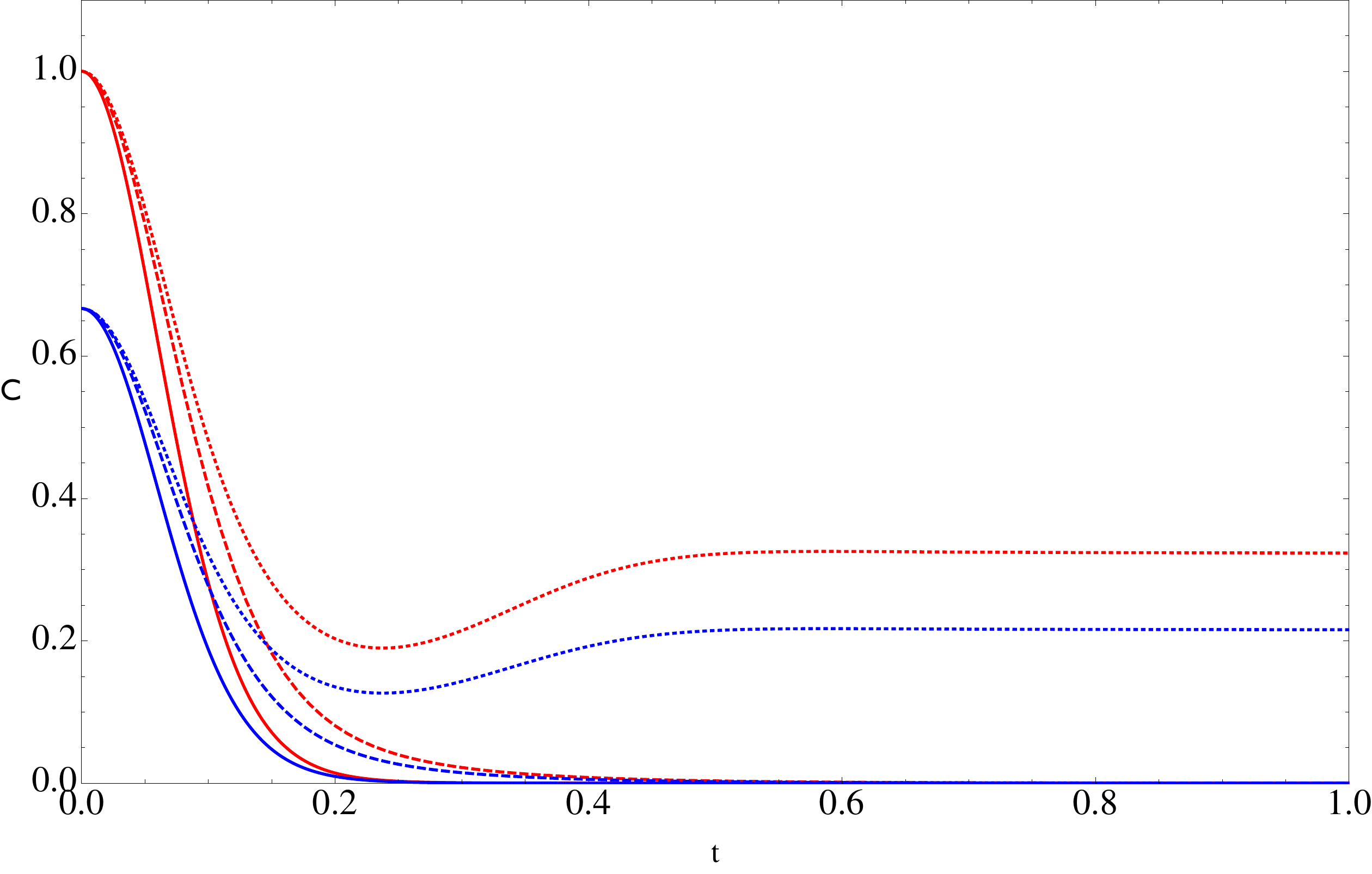}
\caption{Concurrence vs $t$ for the states and environments specified in Fig. \ref{PbNs}.}
\label{CbNs}
\end{figure}

   Now, we consider the decoherence patterns for various initial state $|(e_1,e_2, e_3, e_4) \rangle$'s. As the explicit form of the reduced density matrix is quite tedious and not very illuminating, we directly present the numerical plots. We first consider the uniform environment and plot the decoherence patterns for various initial state $|(e_1,e_2, e_3, e_4) \rangle$.  In each plot we compare the decoherence patterns (either purity or concurrence) for $Q\le 1$ and $Q>1$ cases with the same given initial states.  The results are shown in Fig. \ref{PfEs}-\ref{CbEs}. We then consider some cases for the non-uniform environment, and the results are plotted in Fig. \ref{PfNs}-\ref{CbNs}.

   Moreover, in all the following plots we set $B=0.1$, the cutoff $\Gamma_0=10$,  the parameters $a_0/v$ in \eq{A03} and $\epsilon$ in \eq{B02} to $1/\Gamma_0$ and the parameter $N_{sc}$ in \eq{B02} to one. The unit of time axis is $10/\Gamma_0$. With the above choice of parameters, the fermionic and bosonic environments are in the equal footing so that we can make comparison of the vulnerability of the topological qubits in both environments for a given initial state.

 Fig. \ref{PfEs} is the time evolution of the purity for the uniform fermionic environment with different initial states. Recall that $Q=2\kappa-1$, and indeed the plots for $Q>1$ and $Q \le 1$ behave qualitatively different. For $Q\le 1$ all the initial states decohere into Gibbs state and thermalize as in the case for single topological qubit; for $Q>1$ the initial states first decohere but then regain some coherence, and finally settle down into some states with purity almost equal to one.  This suggests that the super-Ohmic environment with $Q>1$ (or $\kappa>1$, i.e., strongly interacting Luttinger liquids) maintain the purity of the initial states, realizing the robust qubits against quantum decoherence.

 Similarly, as shown in Fig. \ref{CfEs} we find that the patterns of concurrence are different for $Q>1$ and $Q \le 1$. For $Q \le 1$ the concurrence vanishes around the decoherence time scale given in Fig. \ref{PfEs}; for $Q>1$, the concurrence drops a little bit at beginning and then regains into a final value close to the initial one. The non-zero concurrence for the topological qubits in the super-Ohmic environments ensure the state remains quantum even without checking if all the off-diagonal elements of the reduced density matrix diminish or not. Thus, the purity in Fig. \ref{PfEs} for $Q>1$ cases does characterize the quantumness, i.e., how close to the initial pure state.
 
   Note that the concurrence pattern shows a converging behavior, i.e.,  the concurrences of different initial states diminish almost at the same time. This seems to be the case for most of the concurrence plots in this work. However, this in general is not true as shown by the exceptional black solid line in Fig. \ref{CfEs}. Despite that, as compared to the concurrence patterns of the usual non-topological qubits studied in \cite{Hfermion}, the patterns here do have the more converging behaviors.

 The time evolution patterns of the purity and concurrence for the uniform bosonic environments are shown in Fig. \ref{PbEs} and \ref{CbEs}, respectively.  Recall $Q=2\Delta - 4$, in Fig. \ref{PbEs}  and \ref{CbEs} we have plotted the results for both $Q>1$ and $Q\le 1$. Compared to the results in Fig. \ref{PfEs} and \ref{CfEs} for the fermionic environments, we find two different features. The first is as illustrated in \eq{brho-2} and \eq{reduced-D-s}, the purity reduces to the one for the pointer state but not Gibbs state for $Q\le 1$ cases. The second is that the purity and concurrence for $Q>1$ cases though settle down to a final value of non-classical state, which is however quite smaller than the initial one for pure state. This is in contrast to the fermionic case for which the final value is almost the same as the initial one. This is also accompanied by the fact that the decoherence time for the bosonic environments is shorter than the fermionic ones.

  As our choices of parameters are done for putting fermionic and bosonic environments in the equal footing, from the above results we might tempt to say that the fermionic environments are more robust in protecting the quantum informations of the probe topological qubits than the bosonic ones.  This is, in fact, not correct in general. The reason we see the decoherence time for bosonic environment for all the figures we have shown is shorter than fermionic one has to do with the prefactors $c_1(\kappa)$ and $c_2(\Delta)$ of the symmetric Green's functions. The analytic expressions for $c_1(\kappa)$ and $c_2(\Delta)$ are respectively shown in \eq{A04} and \eq{B02-2}. For the same scaling dimension $Q$ ($Q=2\kappa-1$ for fermionic and $Q=2\Delta-4$ for bosonic environment) smaller decoherence time indicates larger prefactor. In our case $c_2((Q+4)/2)>c_1((Q+1)/2)$ for $Q\ge 0.057$ and thus all the results we see seems to suggest the fermionic environment is better in preserving the quantum informations (the smallest $Q$ we show for bosonic case is $0.6$). For non-interacting limit, or $Q=0$, the bosonic environment is actually better in protecting the quantum informations as $c_2(2)<c_1(1/2)$. Thus there is no general rule as to which type of environment is better.

    Here we switch gear to study the effect of the non-uniform environments. As emphasized, this is a peculiar feature for the topological qubits as their component Majorana modes can coupled differently to the environment. The non-uniformity of the environment for the two qubit cases is characterized by the vector of conformal dimensions, i.e., $(\kappa_1,\kappa_2, \kappa_3, \kappa_4)$ (fermionic) and $(\Delta_{12},\Delta_{34})$ (bosonic)  where the sub-indices label the corresponding Majorana modes. Note that for the bosonic cases, two Majorana modes labeled by $a$ and $b$ couple at the same time to an environmental operator of conformal dimension $\Delta_{ab}$.

   Fig. \ref{PfNs} shows the time evolution patterns of purity for non-uniform environments.  A peculiarity in this figure is the behavior of the blue dashed line which does not evolve into Gibbs state as the red dashed line. The blue and red dashed lines are the patterns for the different initial states but in the {\it same} non-uniform environment. Moreover, one can check that the final state for the blue dashed line is not a pointer state, i.e., the reduced density matrix is not a diagonal matrix.  This reflects the fact that the non-uniformity of the environment enhance the breaking of the unitarity of the state space. Thus, some particular subset of states are preferred to be robust against decoherence than the others.
   
       On the other hand, the time evolution pattern of the concurrence shown in Fig. \ref{CfNs} does not have this peculiarity, i.e., the blue dashed line has late-time vanishing concurrence as the red dotted line does. This is because the concurrence characterizes the quantum entanglement between two topological qubits, e.g., the state $e_1|00\rangle + e_4 |11\rangle$ is entangled but $e_1|00\rangle +e_2|01\rangle=|0\rangle(e_1|0\rangle + e_2 |1\rangle$ is not, but both are pure state. Moreover, in our setup specified by \eq{4m-2q} the first topological qubit is made by $\gamma_1$ and $\gamma_2$ and the second one by $\gamma_3$ and $\gamma_4$. Thus, the non-uniform environment $(\kappa_1,\kappa_2,\kappa_3,\kappa_4)=(0.5,0.5,1.2,1.2)$ can retain the quantum information of the second qubit but not the first one. So, this environment may retain the quantumness of the initial state $|(e_1,e_2,e_3,e_4)\rangle=|(2,1,0,2)\rangle$ (the blue dotted line in Fig. \ref{PfNs}, or see Appendix. \ref{A-C1} for more discussions) but not its concurrence as the first topological qubit decoheres into the classical state and can no longer entangle with the second one. Moreover, for both $|(e_1,e_2,e_3,e_4)\rangle=|(2,1,0,2)\rangle$ and $|(1,0,0,1)\rangle$ the ``useful" information counted by concurrence has equal relative weight (i.e. $e_1=e_4$) and coupled to the same environments. Thus for this case the concurrence diminishes exactly at the same time for these two different initial states as shown in dashed and solid lines in Fig. \ref{CfNs}.  
       
       For a general sets of initial states and environment parameters the concurrence needs not diminish at the same time, e.g. the black solid line in Fig. \ref{CfEs} or the blue and red dot-dashed lines in Fig. \ref{CfNs}.  Especially, the latter has the environment with $(\kappa_1,\kappa_2, \kappa_3, \kappa_4)= (0.5,1.2,1.2,0.5)$ which acts on the component Majorana modes of each topological qubit non-uniformly. According to the discussion around \eq{f-nonu-rho}, in this kind of environments  the decoherence pattern of the topological qubit  is very sensitive to the initial states and should also affect the pattern of concurrence between two topological qubits. This may explain why the red and blue dot-dashed lines (the concurrences for initial states $|(e_1,e_2, e_3, e_4) \rangle=|(1,0,0,1)\rangle$ (red), $|(2,1,0,2)\rangle$ (blue) in the environment with $(\kappa_1,\kappa_2, \kappa_3, \kappa_4)= (0.5,1.2,1.2,0.5)$, respectively. ) do not diminish (almost) at the same time  as the other cases in Fig. \ref{CfNs}.

   Similarly, we consider the effect of bosonic non-uniform environments, and the results are shown in Fig. \ref{PbNs} and \ref{CbNs}. Again, as in Fig. \ref{PfNs} we see the similar enhancement effect of the unitarity breaking in Fig. \ref{PbNs}, i.e., the final state purity of the blue dashed line does not merge with the blue solid one which reduces to the pointer state.  One can check the final state of the blue dashed line is not a pointer state. This is in contrast to the merging of the red dashed and solid lines which correspond to another initial state. As for the concurrence, the story is similar to the fermionic case: the environment $(\Delta_{12},\Delta_{34})=(2.3,4.1)$ causes complete decoherence of the first topological qubit so that it cannot maintain the quantum entanglement with the second one for the initial state of the blue dashed line.

\section{Conclusion}\label{secCon}

   In this paper, we investigate the decoherence patterns of topological qubits made of Majorana modes weakly coupled to the Ohmic-like environments so that the topological properties of Kitaev's chain is preserved. Our results give the answers to the following two questions motivated our work.    The first question is whether the topological qubits are robust against decoherence or not. If yes, then the second question is in what sense the robustness means.  For the Ohmic-like environments which are mostly considered in the context of quantum decoherence, we find that the topological qubits cannot completely decohere if the environment is super-Ohmic, so is the concurrence (i.e., quantum entanglement) \cite{fTt}. 

Note that the super-Ohmic environments have less spectral density at low energy than the sub-Ohmic ones. Thus naively we expect the qubits should be more robust against decoherence in the super-Ohmic environment as the low energy degrees of freedom are the main carriers in taking away the quantum information of the probe qubits. Just based on this, however, it is hard to see in what sense the robustness means as there is no real mass gap being developed even for very super-Ohmic spectrum.  In fact, the absence of the mass gap leads to the complete decoherence of the non-topological qubits in all Ohmic-like environments unless the probe-environment coupling is strong enough \cite{Oh,STWu,Hfermion}. 

    Our results clarify the above issues. Even there is no gap developed in the super-Ohmic environment, the influence functional \eq{influence-alpha} does develop an effective mass gap to cutoff the effective carriers so that not all of the quantum information of the topological qubits leaks away. This influence functional is the trademark of topological qubits in contrast to the one for the non-topological qubits. The latter involves also the retarded Green function responsible for the dissipation in the Langevin dynamics of the probe. From our derivation of the open system dynamics, the special form of \eq{influence-alpha} results from the non-local nature of the topological qubits and the peculiar algebra of the Majorana modes. The non-local nature also yields the consideration of the non-uniform environments which cannot be realized for the usual local qubits.  
    
     Despite our results shed new light on the decoherence patterns of the topological qubits in the Ohmic-like environments, we still have not explored all the possible situations. For example, we have not considered the interaction Hamiltonian with non-commuting terms.  Also, one may consider some more general environments other the Ohmic-like cases, such as the one from the holographic gravity duals in the context of AdS/CFT correspondence as proposed in \cite{Ho:2013rra} and studied in \cite{Hfermion}.  It is also interesting to understand the effect of non-uniform environments in a more systematic ways. Finally, to reformulate our derivation of open system dynamics of Majorana modes in the way of Feynman-Vernon will definitely shed new light in understanding our results here in the context of fluctuation and dissipation of the Langevin dynamics.

\acknowledgments
    FLL is supported by Taiwan's NSC grants (grant NO. 100-2811-M-003-011 and 100-2918-I-003-008), and he likes to thank Yong-Shi Wu and Shi Yu for discussions, and Fudan U. and KITPC for hospitality where part of the work was done. All the authors acknowledge the support by NCTS.

\appendix

\section{Schwinger-Keldysh Green functions and KMS condition} \label{A0}
The Schwinger-Keldysh Green functions are the real-time two-point function on the Keldysh contour. They are defined as follows:
\bea\nn \label{app-01}
&&iG^{++}_M(t-t')=\langle \textrm{T} \mathcal{O}_M(t) \mathcal{O}_M(t') \rangle_{\mathcal{E}}\;,\\\nn &&  iG^{--}_M(t-t')=\langle \tilde{\textrm{T}} \mathcal{O}_M(t) \mathcal{O}_M(t') \rangle_{\mathcal{E}}\;,\\\nn
&&iG^{+-}_M(t-t')=\langle \mathcal{O}_M(t) \mathcal{O}_M(t') \rangle_{\mathcal{E}}\;,\\\label{23} && iG^{-+}_M(t-t')=\langle \mathcal{O}_M(t') \mathcal{O}_M(t) \rangle_{\mathcal{E}}\;,
\eea
and their relations to the retarded Green function $G_{R}$ and the symmetric Green function $G_{sym}$:
\bea\label{24}
G_R &=& {1\over 2} (G^{++}-G^{--}-G^{+-}+G^{-+})
\\\label{25}
G_{sym}&=& {i\over 2} (G^{++}+G^{--})={i\over 2}(G^{+-}+G^{-+})\;.
\eea
Moreover, for the thermal environment these two Green functions are related by the Kubo-Martin-Schwinger (KMS) condition \cite{Martin:1959jp}:
\be\label{KMS}
G_{sym}(\omega)=-[1\pm 2 n(\omega)] \textrm{Im} G_R(\omega)
\ee
where the plus sign is for the bosonic channel with $n(\omega)={1\over e^{\beta \omega}-1}$, and the minus sign is for the fermionic channel with $n(\omega)={1\over e^{\beta \omega}+1}$.

\section{On Gaussian approximation of influence functional}\label{reexp}

In this appendix, we compare the explicit expansion in (\ref{CCSS}) and the one by Gaussian approximation (\ref{cc1}), and show that, up to some overall factor in each order, both behave qualitatively the same, i.e., both are just functionals of  $\int^t d\tau \int^t d\tau' \;\overline{G}_{M,sym}(\tau-\tau')$ only. Similar structures are also shown for (\ref{ss1}).

To start, let us expand $\langle \tilde{\textrm{T}} \cosh{\bf O}_M(t) \textrm{T} \cosh{\bf O}_M(t) \rangle_{\mathcal{E}}$ directly:
\beq \nn
&&\langle \tilde{\textrm{T}} \cosh{\bf O}_M(t) \textrm{T} \cosh{\bf O}_M(t) \rangle_{\mathcal{E}}  \\ \nn
&=& \left(1+{x^2 \over2}+{x^4\over 4!}+... \right)\left(1+{y^2 \over2}+{y^4\over 4!}+... \right) \\ \label{AppF01}
&=& 1+ \frac{1}{2}\left( x^2 +y^2 \right)+\frac{1}{4!} \left(x^4+6 x^2 y^2+y^4 \right) +...
\\ \nn
&=& 1+ \frac{1}{4}((x+y)^2+(x-y)^2) 
\\ \nn
&&+\frac{1}{2 \cdot 4!}((x+y)^4+(x-y)^4) + ...
\eeq
where 
\beq
x^2 &\equiv& i \int^t d\tau \int^t d\tau' \;\overline{G}^{++}_M(\tau-\tau'),\nn\\
y^2 &\equiv&i  \int^t d\tau \int^t d\tau' \;\overline{G}^{--}_M(\tau-\tau'),\nn\\
xy &\equiv&i  \int^t d\tau \int^t d\tau' \;\overline{G}^{+-}_M(\tau-\tau')\,\nn\\
yx &\equiv&i \int^t d\tau \int^t d\tau' \;\overline{G}^{-+}_M(\tau-\tau')\nn
\eeq
as  defined in (\ref{app-01}). In the above, the Wick-contraction for higher point points and the symmetrization of $xy$ and $yx$  are assumed. 

On the other hand, if we expand the result based on Gaussian approximation in (\ref{cc1}) in the contour basis, it yields 
\beq \nn
&&\langle \tilde{\textrm{T}} \cosh{\bf O}_M(t) \textrm{T} \cosh{\bf O}_M(t) \rangle_{\mathcal{E}}  \\ \nonumber
&\approx&{1\over 2}(e^{(x+y)^2/2}+e^{(x-y)^2/2})  \\ \label{AppF02}
&=& 1 + \frac{1}{2} \left( x^2 +y^2 \right) + \frac{1}{8} \left(x^4+6 x^2 y^2+y^4 \right) +...\;.
\eeq

Comparing (\ref{AppF01}) and (\ref{AppF02}), we can see that these two expansions differ by a factor of $(2n-1)!!$ for nth order terms, i.e., in some sense the exact correlator and its Gaussian approximation are Borel-like-sum related in the expansion of $(x-y)^2$ and $(x+y)^2$. Note that if we do the Gaussian approximation in the other way, e.g., $\frac{1}{2}(e^{x^2}+e^{y^2})$, it will not be Borel-like-sum related to \eq{AppF01} though the first order terms are the same, and will yield incorrect long time behaviors \cite{reexp-fn}.

We should emphasize that this kind of Gaussian approximation is usually adopted in the Feynman-Vernon's way of deriving the quadratic form of influence functional, e.g. see \cite{Hu:1991di,Boyanovsky:2004dj}. The difference here is that we have first eliminated the probe field $\gamma$'s by using $\gamma^2=1$. Otherwise, the algebraic structures for ours and the Feynman-Vernon one are the same.

The above demonstration shows that, after transforming back to the ``ra" basis, both expressions are order by order just functionals of $\int^t d\tau \int^t d\tau' \;\overline{G}_{M,sym}(\tau-\tau')$, i.e., $(x+y)^2 \equiv 4  \int^t d\tau \int^t d\tau' \;\overline{G}_{M,sym}(\tau-\tau')$ and $(x-y)^2 \equiv 0$ by the second equality of \eq{25}.  This indicates that the critical behaviour occurred at $Q=1$ is qualitatively true to all orders as the criticality is just  encoded in $\int^t d\tau \int^t d\tau' \;\overline{G}_{M,sym}(\tau-\tau')$. Thus, the Gaussian approximation used in this work still ensures the qualitative difference between the sub- and super-Ohmic environments when considering the decoherence patterns of the topological qubits. On the other hand, the quantitative difference is the same as the usual case of Gaussian approximation in the Feynman-Vernon approach, and can be seen as the appropriate mean field approximation.

\section{Green functions of the Fermionic and Bosonic environments}

\subsection{Fermionic Green function}\label{A-1}

Here we list the Green's function for the fermionic environment considered in the main text. Following \cite{SungPo1}, the Keldysh component of bare (uncoupled) zero temperature helical Luttinger liquids lead Green functions in frequency space are:
\bes \label{A01}
\beq\label{A01-1}
G_{\psi_{L/R}}^{++}(\omega)&=&\frac{a_0^{2\kappa}}{2\pi v^{2 \kappa}} \frac{\Gamma(\kappa)^2}{\Gamma(2\kappa)} |\omega-\mu|^{2\kappa-1}\\\nn&&  \left( \tilde{h}(\kappa) \theta(\omega-\mu)-\tilde{h}(\kappa) \theta(\mu-\omega)\right) \\\label{A01-2}
G_{\psi_{L/R}}^{--}(\omega)&=&\frac{a_0^{2\kappa}}{2\pi v^{2 \kappa}} \frac{\Gamma(\kappa)^2}{\Gamma(2\kappa)} |\omega-\mu|^{2\kappa-1}\\\nn &&  \left( \tilde{h}^*(\kappa) \theta(\mu-\omega) -\tilde{h}^*(\kappa) \theta(\omega-\mu)\right) \\
 \label{A01-3}
G_{\psi_{L/R}}^{+-}(\omega)&=&\frac{2 \pi a_0^{2\kappa}}{ v^{2 \kappa}} \frac{ i }{\Gamma(2\kappa)} |\omega-\mu|^{2\kappa-1} \theta(\mu-\omega) \\
 \label{A01-4}
G_{\psi_{L/R}}^{-+}(\omega)&=&\frac{2 \pi a_0^{2\kappa}}{ v^{2 \kappa}} \frac{ -i }{\Gamma(2\kappa)} |\omega-\mu|^{2\kappa-1} \theta(\omega-\mu)
\eeq
\ees
where 
$\kappa=\frac{1}{4} (K+1/K)$ and $\tilde{h}(\kappa)=2 e^{-i\pi \kappa} sin(\pi\kappa) \Gamma(1-\kappa)^2$. $K=1$ ($\kappa=1/2$) corresponds to non-interacting fermions and L/R indices correspond to left/right movers in the one dimensional system. Notice the relative minus sign in \eq{A01-1} and \eq{A01-2} due to their fermionic nature.

For our purpose, we have to modify the Green functions in (\ref{A01}) to ``Majorana-dressed" ones. They are
\bes \label{A02}
\beq \label{A02-1}
\overline{G}_{\psi_{L/R}}^{++}(\omega)&=&\frac{a_0^{2\kappa}}{2\pi v^{2 \kappa}} \frac{\Gamma(\kappa)^2}{\Gamma(2\kappa)} |\omega-\mu|^{2\kappa-1}\\\nn && \left( \tilde{h}(\kappa) \theta(\omega-\mu)+\tilde{h}(\kappa) \theta(\mu-\omega)\right) \\
 \label{A02-2}
\overline{G}_{\psi_{L/R}}^{--}(\omega)&=&\frac{a_0^{2\kappa}}{2\pi v^{2 \kappa}} \frac{\Gamma(\kappa)^2}{\Gamma(2\kappa)} |\omega-\mu|^{2\kappa-1}\\\nn &&\left( -\tilde{h}^*(\kappa) \theta(\omega-\mu)-\tilde{h}^*(\kappa) \theta(\mu-\omega)\right) \\
 \label{A02-3}
\overline{G}_{\psi_{L/R}}^{+-}(\omega)&=&\frac{2 \pi a_0^{2\kappa}}{ v^{2 \kappa}} \frac{ -i }{\Gamma(2\kappa)} |\omega-\mu|^{2\kappa-1} \theta(\mu-\omega) \\
 \label{A02-4}
\overline{G}_{\psi_{L/R}}^{-+}(\omega)&=&\frac{2 \pi a_0^{2\kappa}}{ v^{2 \kappa}} \frac{ -i }{\Gamma(2\kappa)} |\omega-\mu|^{2\kappa-1} \theta(\omega-\mu)
\eeq
\ees
It is easy to show the relation between contour Green function $\overline{G}^{++}+\overline{G}^{--}=\overline{G}^{+-}+\overline{G}^{-+}$ still holds for these ``Majorana-dressed" Green functions.

   Note that in our context we assume the Majorana zero modes coupled locally to the anti-Hermitian operator $\mathcal{O}_a$ as noted in \eq{antihermi}. Here we assume $\mathcal{O}_a=\psi_{L/R}-\psi^{\dagger}_{L/R}$, which is anti-hermitian by construction. 
Then our symmetric Green function for operator $\mathcal{O}$ is 
\beq \nn
\overline{G}_{\mathcal{O}_a,sym}(\omega)&=&\frac{i}{2} ( \overline{G}^{++}_{\mathcal{O}_a}(\omega) +  \overline{G}^{++}_{\mathcal{O}_a}(\omega)) \\
&=& -i  (\overline{G}^{++}_{\psi_{L/R}}(\omega)+\overline{G}^{--}_{\psi_{L/R}}(\omega) )  \nonumber \\
&=&- \frac{a_0^{2\kappa}}{\pi v^{2 \kappa}} \frac{\Gamma(\kappa)^2}{\Gamma(2\kappa)} ( 2 i \mbox{Im} \tilde{h}(\kappa) ) |\omega-\mu|^{2\kappa-1} \nonumber \\
&=&c_1(\kappa) |\omega-\mu|^{2\kappa-1}
\eeq
Here we have used Euler's reflection formula $\Gamma(\kappa)^2 \Gamma(1-\kappa)^2 \mbox{sin}^2\pi\kappa =\pi^2$ from second to third line and defined $c_1(\kappa)$ as:
\be\label{A04}
c_1(\kappa) \equiv -\frac{4 \pi}{\Gamma(2\kappa)} \left(\frac{a_0}{v} \right)^{2\kappa}<0.
\ee

\subsection{Bosonic Green function}\label{A-2}

In bosonic case, we consider the Green function of a scalar field in $AdS_5$ space in long wavelength ($|\vec{k}| \sim 0$) limit such that its Green function effectively reduces to zero dimension \cite{Son:2002sd,Ho:2013rra}. We denote the dual operator of the scalar by $\mathcal{O}_s$, then its holographic retarded Green function is
\begin{widetext}
\begin{equation}\label{B01}
G_{\mathcal{O}_s,\textrm{R}}(\omega)=\begin{cases}
\frac{N_{sc}^2 \Gamma(3-\Delta) \epsilon^{2(\Delta-4)}}
{8 \pi^2 \Gamma(\Delta-2) 2^{2\Delta-5}} \; 
(\omega^2)^{\Delta-2}\; [\; \cos \pi \Delta - i\;  \textrm{sgn}(\omega) 
\sin \pi\Delta\; ]& 2<\Delta\notin \mathbb{N}\\
\frac{N_{sc}^2 \epsilon^{2(\Delta-4)}}{8\pi^2 
( \Delta-3)!^2 2^{2\Delta-5}}\; (\omega^2)^{\Delta-2}\; [\; 
\ln \omega^2 - i\;\pi\; \textrm{sgn}(\omega)\;]& 2\leq\Delta\in \mathbb{N}
\end{cases}
\end{equation}
\end{widetext}
where $N_{sc}^2$ is the number of degrees of freedom of the dual conformal field theory, and $\epsilon \approx 0$   is the UV cutoff of length scale.

   Simialr to the fermionic case, the bosonic operator $\mathcal{O}_{ab}$ to which the double Majorana modes couple is anti-hermitian as noted in \eq{antihermi}.  We then assume $\mathcal{O}_{ab}=\mathcal{O}_s-\mathcal{O}^{\dagger}_s$ so that $G_{\mathcal{O}_{ab},R}=-2 G_{\mathcal{O}_s,R}$. Thus,  the symmetric Green function of $\mathcal{O}_{ab}$ related to the retarded one (at zero temperature) by \eq{KMS} is given
\bes \label{B02}
\beq \nn
G_{\mathcal{O}_{ab}, sym}(\omega) &=& 2 \mbox{ sgn}(\omega) \mbox{ Im} G_{\mathcal{O}_s,R}(\omega)\\\label{B02-1}
&=&c_2(\Delta) (\omega^2)^{\Delta-2}
\eeq
where 
\begin{equation}\label{B02-2}
c_2(\Delta) \equiv \begin{cases}
-\frac{N_{sc}^2 \Gamma(3-\Delta) \epsilon^{2(\Delta-4)}}
{4 \pi^2 \Gamma(\Delta-2) 2^{2\Delta-5}} \; 
(\sin \pi\Delta)\; & 2<\Delta\notin \mathbb{N}\\
-\frac{N_{sc}^2 \epsilon^{2(\Delta-4)}}{4\pi 
( \Delta-3)!^2 2^{2\Delta-5}}\;  & 2\leq\Delta\in \mathbb{N}
\end{cases}
\end{equation}
\ees
Note that for bosonic channel, $\overline{G}_{sym}(\omega)=G_{sym}(\omega)$.

\section{Purity for single qubit system}\label{A-B2}

In this section, we record the explicit form of purity for single qubit system. The explicit form of the reduced density matrix for fermionic and bosonic environments are given in 
Eq. (\ref{frho-1}) and Eq. (\ref{brho-1}) respectively. It is then straightforward to calculate the purity $\mathcal{P}(t) \equiv \mbox{Tr } \rho(t)^2$. The explicit forms are given below:

\begin{widetext}
For bosonic environment, the purity takes the form
\beq \label{purity-single-b}
\mathcal{P}^b(t) = \frac{1}{N^b(t)^2} \left(a_{01}a_{10} (C_{12}+S_{12})^2+(a_{00}^2+a_{11}^2)(C_{12}-S_{12})^2\right)
\eeq
and 
\beq \label{purity-single-f}
\mathcal{P}^f(t) &&= \frac{1}{N^f(t)^2} \left( 2 (a_{01} (C_1 C_2-S_1 S_2)+a_{10} (C_1 S_2 - C_2 S_1) ) ( a_{10} (C_1 C_2 - S_1 S_2)+a_{01} (C_1 S_2- C_2 S_1)) \right) \nonumber \\
&&+\frac{1}{N^f(t)^2} \left( (a_{11}(C_2 S_1+C_1 S_2)-a_{00} (C_1 C_2 + S_1 S_2) )^2 + (a_{00} (C_2 S_1 +C_1 S_2) - a_{11} (C_1 C_2 +S_1 S_2))^2     \right)
\eeq
for fermionic environment. Here $N^{f,b}(t)=\Tr \; \rho^{f,b}_r(t)$.
\end{widetext}

\section{Reduced density matrix of two topological qubits}\label{A-B1}

Here we write down the explicit form for the reduced density matrix of two topological qubits with 
the initial states as a superposition of even fermion parity states, i.e. $\psi(t=0) = a |00 \rangle +b |11 \rangle$ with $|a|^2+|b|^2=1$. Therefore, the initial density matrix (in qubit basis $(|00 \rangle,\ |01\rangle,\ |10\rangle,\ |11\rangle)^T$) is given in \eq{2qbell}.

Using \eq{rhofinal-s}, the non-vanishing reduced density matrix elements $\rho_{r,ij}^f(t)$ for fermionic environmental channel at time $t$ are
\begin{widetext}
\bes \label{D03}
\beq 
\label{D03-1}
\mathcal{N}^f(t) \rho^f_{r,11}(t)&& =(C_1 C_2 + S_1 S_2) (C_3 C_4 + S_3 S_4) |a|^2 + (C_2 S_1 + C_1 S_2) (C_4 S_3 + C_3 S_4) |b|^2 \;,\\
\label{D03-2}
\mathcal{N}^f(t) \rho^f_{r,14}(t)&&= (C_1 C_2 - S_1 S_2) (C_3 C_4 - S_3 S_4)ab^* - (C_2 S_1 - C_1 S_2) (C_4 S_3 - C_3 S_4) a^* b=\mathcal{N}(t) (\rho^f_{41}(t) )^*\;,\\
\label{D03-3}
\mathcal{N}^f(t) \rho^f_{r,22}(t)&& =-(C_1 C_2 + S_1 S_2) (C_4 S_3 + C_3 S_4)  |a|^2  - (C_2 S_1 + C_1 S_2) (C_3 C_4 + S_3 S_4) |b|^2\;, \\
\label{D03-4}
\mathcal{N}^f(t) \rho^f_{r,23}(t)&&= (C_1 C_2 - S_1 S_2) (C_4 S_3 - C_3 S_4)ab^* - (C_2 S_1 - C_1 S_2) (C_3 C_4 - S_3 S_4)a^* b=\mathcal{N}(t) (\rho^f_{32}(t))^*\;,\\
\label{D03-5}
\mathcal{N}^f(t) \rho^f_{r,33}(t)&& =-(C_2 S_1 + C_1 S_2) (C_3 C_4 + S_3 S_4)  |a|^2 -(C_1 C_2 + S_1 S_2) (C_4 S_3 + C_3 S_4)|b|^2\;, \\
\label{D03-6}
\mathcal{N}^f(t) \rho^f_{r,44}(t)&& =(C_2 S_1 + C_1 S_2) (C_4 S_3 + C_3 S_4)   |a|^2 +(C_1 C_2 + S_1 S_2) (C_3 C_4 + S_3 S_4)|b|^2\;, \\
\label{D03-7} 
\mathcal{N}^f(t) && = (C_1 - S_1) (C_2 - S_2) (C_3 - S_3) (C_4 - S_4) (|a|^2 +|b|^2 )\;.
\eeq
\ees
\end{widetext}

Similarly, the non-vanishing $\rho_{r,ij}^b$ for bosonic channel are

\bes \label{D04}
\beq
\label{D04-1}
 \rho^b_{r,11}(t) &= &{1\over \mathcal{N}^b(t)} |a|^2 (C_{12} - S_{12}) (C_{34} - S_{34}), \\
\label{D04-2}
 \rho^b_{r,44}(t) &=& {1\over \mathcal{N}^b(t)} |b|^2 (C_{12} - S_{12}) (C_{34} - S_{34}), \\
\label{D04-3}
\rho^b_{r,14}(t)&=& {1\over \mathcal{N}^b(t)} a^*b  (C_{12} +S_{12}) (C_{34} +S_{34}), \\
\label{D04-4}
\mathcal{N}^b(t)  &=&  (|a|^2+|b|^2) (C_{12} - S_{12}) (C_{34} - S_{34}) 
\eeq
\ees
with $(\rho^b_{41}(t))=(\rho^b_{14}(t))^*$.
Note that in this case we only turn on the parity-conserving interactions $\gamma_1\gamma_2 \mathcal{O}_{12}$ and $\gamma_3\gamma_4 \mathcal{O}_{34}$.

\section{Two topological qubits with nonuniform fermionic environment}\label{A-C1}

Here we discuss in more details about the effect of nonuniform fermionic environments on two topological qubits as in Fig.\ref{PfNs} and Fig.\ref{CfNs}. We choose $(\kappa_1,\kappa_2,\kappa_3,\kappa_4)=(0.5,0.5,\kappa_0,\kappa_0)$ with $\kappa_0\gg 1$ (indicating $\gamma_3$ and $\gamma_4$ does not leak information into the environments) and the two qubits bases chosen as in the main text. To illustrate the physics we compare the long time behavior under this nonuniform environments with two initial states given by $(e_1,e_2,e_3,e_4)=(1,0,0,1)$ and $(e_1,e_2,e_3,e_4)=(1,1,0,0)$. 

 For the second choice the initial state is in the form of a product state ($|00\rangle+|01\rangle$) and the concurrence is zero for all time. For the qubits bases $|i j\rangle$ we expect the information contained in $``j"$ qubit does not leak into the environment as $\kappa_0\gg 1$, but the the information contained in $``i"$ qubit can leak to the environments as $\gamma_1$ and $\gamma_2$ are connected with Fermi liquids leads. At long time from statistical argument we expect the probability for $i=0$ and $i=1$ should be equal as these two states are energetically degenerate. Thus for $(e_1,e_2,e_3,e_4)=(1,0,0,1)$ or initial probability $1/2$ at $|00\rangle$ and $1/2$ at $|11\rangle$, we expect the $|00\rangle$ state becomes $|00\rangle$ and $|10\rangle$ with probability $1/4$ and   
the $|11\rangle$ state becomes $|01\rangle$ and $|11\rangle$ with probability $1/4$. Since the quantum information is carried by superposition of $|00\rangle$ and 
$|11\rangle$, dephasing at first qubit erases this mutual information/memory completely and the reduced density matrix $\rho^{f(1)}_r(t)$ at late time becomes

\beq \label{reduced-Df-exp1} 
\rho^{f(1)}_r(t\rightarrow\infty)=\left(\begin{array}{cccc}\frac{1}{4} & 0 & 0 & 0 \\0 & \frac{1}{4}  & 0 & 0 \\0 & 0 & \frac{1}{4}  & 0 \\ 0 & 0 & 0 & \frac{1}{4} \end{array}\right)
\eeq
This is not the case for the second choice of initial state $(e_1,e_2,e_3,e_4)=(1,1,0,0)$ as shown below. Following the same argument as in $(1,0,0,1)$, we expect the $|00\rangle$ state becomes $|00\rangle$ and $|10\rangle$ with probability $1/4$ and the $|01\rangle$ state becomes $|01\rangle$ and $|11\rangle$ with probability $1/4$ at late time. The key difference from the $(1,0,0,1)$ case is that the relative phase information for $(1,1,0,0)$ is stored at second qubit \emph{only} and the random fluctuations from the first qubit does not influence it. Mathematically speaking, $(1,1,0,0)$ initial state is $|00\rangle+|01\rangle=|0\rangle\otimes(|0\rangle+|1\rangle)$, and as such the random change on the first qubit ($``|0\rangle"$ state) does not alter the relative phase information $(|0\rangle+|1\rangle)$ of the second qubit. The reduced density matrix $\rho^{f(2)}_r(t)$ at late time becomes  
\beq \label{reduced-Df-exp2} 
\rho^{f(2)}_r(t\rightarrow\infty)=\left(\begin{array}{cccc}\frac{1}{4} & \frac{1}{4} & 0 & 0 \\\frac{1}{4} & \frac{1}{4}  & 0 & 0 \\0 & 0 & \frac{1}{4}  & \frac{1}{4}  \\ 0 & 0 & \frac{1}{4}  & \frac{1}{4} \end{array}\right)
\eeq
 From \eq{reduced-Df-exp2} we see the reduced density matrix does not go to pointer state with $(e_1,e_2,e_3,e_4)=(1,1,0,0)$ but it does go to Gibbs state for $(1,0,0,1)$ in this case. The physics described in this example explains why we see those different behaviors in Fig.\ref{PfNs} and Fig.\ref{CfNs}, and for nonuniform with some super-Ohmic environment(s) even the fermionic channels does not necessarily go to Gibbs state.


\begin{thebibliography}{99}
\bibitem{finiteT}
For example, in the real work the anyons are in contact with a thermal bath. In this case, the robustness of topological order has to be reconsidered. For instance, there are results indicating the lack of robustness of topological order in 2D cases \cite{FiniteT-1} while in higher dimensions there exists robust proposal for self-correcting computation \cite{FiniteT-2}. However, we restrict ourselves only for zero temperature cases in this work.
\bibitem{superselection}
A single topological qubit considered in \cite{Chamon,Budich,Loss} is made by the parity-even sector of 4 Majorana modes, instead of 2 Majorana modes considered in this work. This is because, by the super-selection rule one cannot observe relative phase in the linear superposition of two degenerate eigenstates of a parity operator. In this work, we by-pass the super-selection rule by preparing the general (parity-broken) initial state induced by the parity-violating operator. This can be done either by shortening the Kitaev's chain (so that the tunneling strength is increasing) or by turning on the parity-breaking probe-environment operators (recall the distance between 2 Majorana modes of a single qubit, especially in ring case, is very small in our setup.). Once the general initial states are prepared, our dynamics follows.  On the other hand, qubit made of even sector of 4 Majorana modes will be a special case of our two qubits' cases so that our conclusion also holds in such a reduced case.
\bibitem{reexp-fn}
The leading order after expanding $\cosh O_M$ and $\sinh O_M$ is $\mathcal{O}(B_M^2)$ as $\mathcal{O}(B_M)$ has zero expectation value from \eq{nofvev}. We then re-exponentiate the leading $\mathcal{O}(B_M^2)$ terms without regrouping terms to avoid nontrivial cancellations and express the various propagators in the exponent via \eq{23}-\eq{25}.
\bibitem{parity-issue}
Another way to see this is through the counting of Majorana fermions involved in the coupling terms. For odd number of Majorana fermions (generic fermionic environments) the even and odd fermion initial states could mix through this interaction, leading to thermailzed states as expected. While for even number of Majorana fermions (bosonic environments) the even/odd fermion states could only mix with even/odd states. Thus the bosonic environments bring in this additional symmetry which prevents the thermalization in the sub-Ohmic environments. Parity-preserving coupling, which is the same as Majorana fermion number counting for single qubit but different for two qubits case, imposes even stronger constraint such that the even/odd state do not mix with another even/odd states, maintaining the diagonal components intact  at long time limit. For super-Ohmic environments both fermionic and bosonic environments do not decohere and thus do not reach pointer states.
\bibitem{fTt}
A type of partial preservation of the topological structure at finite temperature has been reported for topological insulators in \cite{Lindblad-2}.
\end{thebibliography}
\end{document}